\documentclass[pdflatex]{sn-jnl}

\usepackage[T1]{fontenc}
\usepackage[utf8]{inputenc}
\usepackage{amsmath,amsthm}
\usepackage{newtxtext,newtxmath}
\usepackage{microtype}
\usepackage{bm}
\usepackage{graphicx,booktabs,array,longtable}
\usepackage{xcolor}
\usepackage{placeins}
\usepackage{bookmark}
\hypersetup{colorlinks=true,linkcolor=black,citecolor=black,urlcolor=blue!55!black}
\usepackage[super,sort&compress]{natbib}
\newcommand{\ket}[1]{\lvert #1\rangle}
\newcommand{\bra}[1]{\langle #1\rvert}
\newcommand{\tr}{\operatorname{Tr}}
\newcommand{\id}{\mathbb{I}}
\newcommand{\Cstate}{\ket{C_N}}
\newcommand{\Heff}{H_{\mathrm{eff}}}
\newcommand{\Fid}{F_C}
\theoremstyle{thmstyleone}

\graphicspath{{../figures/}{figures/}}
\begin{document}
\title{Nondemolition filtering of an embedded cluster-state scar under continuous local monitoring}

% 第一作者：王熙墨
\author[1,2]{\fnm{Ximo} \sur{Wang}}
% 第二作者：赵茜
\author[3]{\fnm{Xi} \sur{Zhao}}
% 第三作者：孙夏宇
\author[1,2]{\fnm{Xiayu} \sur{Sun}}
\author[1,2]{\fnm{Qiwei} \sur{Han}}

\author[4]{\fnm{Yuhang} \sur{Wang}}

% 第六作者：Chunxiao Du
\author[5]{\fnm{Chunxiao} \sur{Du}}

% 第九作者、通讯作者：李睿
\author*[6]{\fnm{Rui} \sur{Li}}
\email{rli.work@buaa.edu.cn}

% 第七作者：Wenxiu Li
\author[7]{\fnm{Wenxiu} \sur{Li}}

% 第八作者：张浩
\author[8]{\fnm{Hao} \sur{Zhang}}

% 单位 1
\affil[1]{
	\orgdiv{School of Physics and Electronic Engineering},
	\orgname{Shanxi University},
	\orgaddress{\city{Taiyuan}, \country{China}}}

% 单位 2：沿用上一版的极端光学协同创新中心英文名称
\affil[2]{
	\orgdiv{Collaborative Innovation Center of Extreme Optics},
	\orgname{Shanxi University},
	\orgaddress{\city{Taiyuan}, \country{China}}}

% 单位 3
\affil[3]{
	\orgdiv{Department of Physics},
	\orgname{University of Science and Technology of China},
	\orgaddress{\city{Hefei}, \country{China}}}

% 单位 4
\affil[4]{
	\orgdiv{School of Instrument Science and Opto-Electronics Engineering},
	\orgname{Beijing Information Science and Technology University},
	\orgaddress{\city{Beijing}, \country{China}}}

% 单位 5
\affil[5]{
	\orgdiv{School of Physics},
	\orgname{Beihang University},
	\orgaddress{\city{Beijing}, \country{China}}}

% 单位 6
\affil[6]{
	\orgdiv{School of Applied Science},
	\orgname{Beijing Information Science and Technology University},
	\orgaddress{\city{Beijing}, \country{China}}}
% 单位 7
\affil[7]{
	\orgdiv{School of Automation (School of Artificial Intelligence)},
	\orgname{Beijing Information Science and Technology University},
	\orgaddress{\city{Beijing}, \country{China}}}

% 单位 8
\affil[8]{
	\orgdiv{School of Space and Earth Sciences},
	\orgname{Beihang University},
	\orgaddress{\city{Beijing}, \country{China}}}

\abstract{
Identifying a low-entanglement eigenstate inside a many-body spectrum and preserving it during measurement are distinct tasks. We construct an explicit local ring Hamiltonian with an exact cluster-state eigenvector and study continuous monitoring of its stabilizer defects. For arbitrary mixed inputs, the conditional cluster fidelity is the initial target weight divided by the no-observed-click probability. A positive defect-operator gap gives finite-time bounds that hold for noncommuting Hamiltonian dynamics, nonnormal effective generators and imperfect detection. At fixed total monitoring rate, the guaranteed exponent falls inversely with system size; high conditional fidelity does not remove the preparation cost set by the initial overlap. Exact diagonalization up to eleven qubits gives finite-size evidence for a cluster-state outlier in a chaotic spectral background (adjacent-gap ratio $\langle r \rangle \approx 0.596$). Independent matrix and trajectory calculations verify the dynamics and a conservative coherent-error bound. This construction specializes established scar embedding and nondemolition verification frameworks, with explicit measurement assumptions, finite-time guarantees and resource limitations, supported by Lean 4 formal verification.
}

\maketitle

\section{Introduction}
The eigenstate thermalization hypothesis connects microscopic quantum dynamics with equilibrium statistical mechanics.\cite{Deutsch1991,Srednicki1994,Rigol2008,DAlessio2016,Mori2018} Quantum many-body scars are exceptional nonthermal eigenstates embedded in otherwise thermalizing systems; a specially arranged tower can also generate coherent revivals.\cite{Bernien2017,Turner2018,Turner2018PRB,Serbyn2021,Moudgalya2022,Chandran2023} Exact embedding methods show that these exceptional states can coexist with local interacting Hamiltonians.\cite{ShiraishiMori2017,LinMotrunich2019,Moudgalya2018,SchecterIadecola2019,Mark2020} An exact eigenvector alone does not establish a chaotic background, and neither property determines how an actual detector changes the state.

Cluster states provide a useful setting because their entanglement and local stabilizers are exactly characterized.\cite{BriegelRaussendorf2001,RaussendorfBriegel2001,Raussendorf2003,Hein2004} Stabilizer scars, cluster-state parent Hamiltonians and single-scar observation protocols have already been developed.\cite{StabilizerScars2025,Dooley2026,Larsen2026} Scalable verification of scar-based quantum simulations has also been proposed.\cite{Hartse2026Benchmark} These developments motivate an operational question: what can a local monitoring record certify while a noncommuting many-body Hamiltonian remains active?

Local correlation fingerprints, exemplified by Li and colleagues,\cite{Li2026Fingerprint} give an economical representation of structured states. A classifier trained on a promised ensemble, however, differs from an unrestricted quantum-state certificate. A single fixed product-basis distribution always has a fully separable explanation: completely dephasing the input in that basis leaves every measured probability unchanged. Measuring each Pauli factor separately also resolves information that a stabilizer parity measurement leaves unresolved. The resulting instruments produce different disturbance even when their classical outcomes estimate the same correlation. We therefore move from correlation readout to ancilla-assisted defect monitoring.

Nondemolition entanglement verification, including stabilizer states, is established,\cite{Liu2021QND,Dangniam2020,Riera2023} as are decoherence-free scar embeddings and non-Hermitian scar stabilization.\cite{Wang2024DFS,Chen2023NH,Omiya2025} We do not claim the first cluster scar, the first nondemolition stabilizer test, or a new non-Hermitian phase. We develop a fully specified benchmark that connects an exact local embedding to mixed-state, finite-efficiency filtering inequalities and explicit resource accounting. The analytic results hold at arbitrary system size; evidence for a chaotic bulk remains limited to the sizes diagonalized.

\section{Results}
\subsection{A local Hamiltonian with an embedded cluster eigenstate}
Consider an $N\geq5$ qubit ring, with indices understood modulo $N$. Let $\mathsf h=(X+Z)/\sqrt2$ be the one-qubit Hadamard gate and define
\begin{equation}
W=\left(\prod_{i=0}^{N-1}\mathrm{CZ}_{i,i+1}\right)\mathsf h^{\otimes N},\qquad
\Cstate=W\ket{0}^{\otimes N},\qquad K_i=Z_{i-1}X_iZ_{i+1}.
\label{eq:cluster}
\end{equation}
The independent commuting stabilizers $K_i$ define the target projector $P=\Cstate\bra{C_N}$ and defect projectors $Q_i=(\id-K_i)/2$. We write $R=\id-P$. In the syndrome frame, $n_i=(\id-Z_i)/2$ and $Wn_iW^\dagger=Q_i$.

An explicit family is
\begin{equation}
\widetilde H=J\sum_{i=0}^{N-1}\sum_{\delta=\pm1}n_i
\left(a_{i\delta}X_{i+\delta}+b_{i\delta}Y_{i+\delta}+c_{i\delta}Z_{i+\delta}\right),
\qquad H=W\widetilde H W^\dagger .
\label{eq:model}
\end{equation}
All coefficients are real. Each control commutes with its neighbouring Pauli operator, so every term is Hermitian. The relations $\widetilde H\ket{0}^{\otimes N}=0$ and $H\Cstate=0$ hold exactly. Conjugation gives $WX_jW^\dagger=Z_j$, $WY_jW^\dagger=-Z_{j-1}Y_jZ_{j+1}$ and $WZ_jW^\dagger=K_j$; each term of $H$ therefore acts on at most four consecutive ring sites. The construction specializes annihilator-based embedding and introduces no new embedding principle.\cite{ShiraishiMori2017,Moudgalya2020MPS,Dooley2026}

For the numerical ensemble, the $6N$ coefficients are sampled independently and uniformly on $[-1,1]$, using prescribed seeds with no selection by the observed spectrum. The computational-basis transition graph contains an isolated vacuum and, almost surely, one connected component containing all nonempty bitstrings. Any nonempty configuration can grow to the fully occupied ring by facilitated flips. After graph dressing, the algebra generated by the independent local terms has commutant $\operatorname{span}\{P,R\}$; Supplementary Note~2 gives a proof. This rules out additional coefficient-independent fragmented components for the whole term family. It does not establish thermalization of every fixed coefficient realization.

\begin{figure}[htbp]
\centering\includegraphics[width=\linewidth]{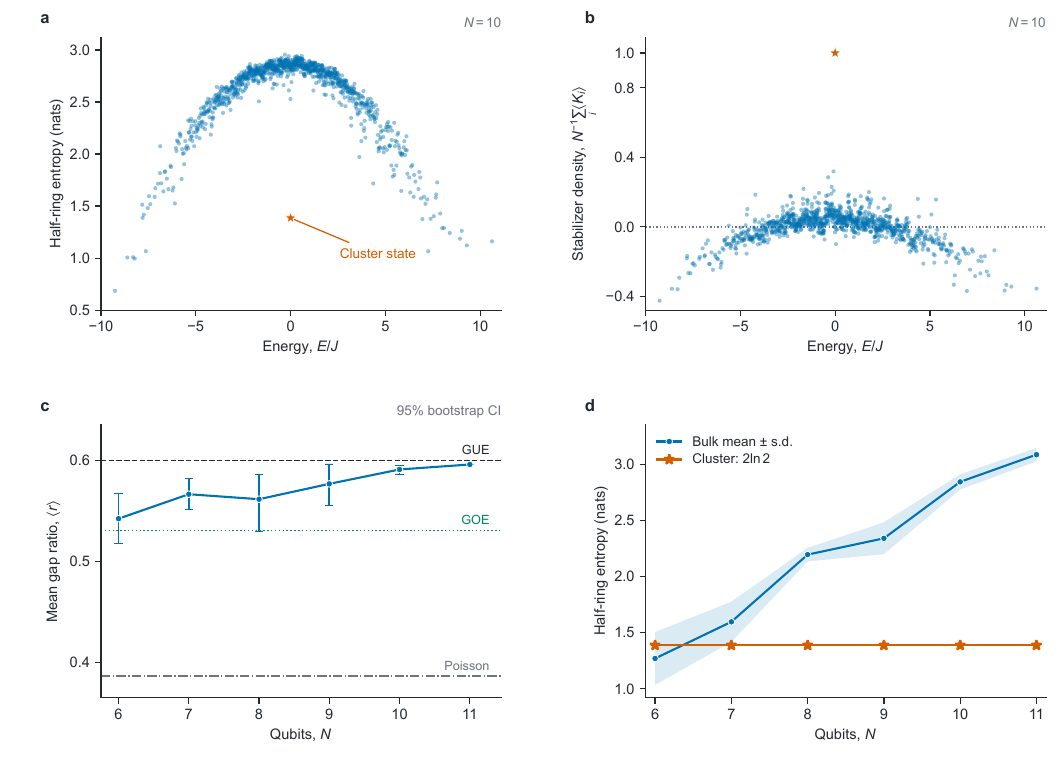}
\caption{\textbf{An exact cluster eigenstate and finite-size bulk diagnostics.} \textbf{a,b}, Half-ring entropy and mean stabilizer expectation for every complementary eigenstate of the $N=10$ representative Hamiltonian, together with the exact target (star). Energies use $J=1$. \textbf{c}, Adjacent-gap ratios in the central half of the complementary spectrum, averaged over $12,12,10,8,6,3$ prescribed realizations for $N=6,\ldots,11$. Error bars are percentile bootstrap intervals over realization means, not a thermodynamic uncertainty estimate; the last size has only three realizations. \textbf{d}, Exact target entropy and mean entropy of up to 64 complementary states nearest zero, using the first realization at each size. Shading denotes the standard deviation across those eigenstates, not disorder-averaged uncertainty.}
\label{fig:embedding}
\end{figure}

Figure~\ref{fig:embedding} separates the exact construction from the finite-size evidence. For the representative $N=10$ sample, the cluster eigenstate lies at the $50.54\%$ energy percentile of the complementary spectrum. Its half-ring entropy is exactly $2\ln2$; the 64 complementary eigenstates closest to zero energy have mean entropy $2.842$ nats and entropy spread $0.068$ nats. The mean stabilizer density $N^{-1}\sum_iK_i$ equals one in the target and lies near zero in the surrounding eigenstates. Since the trace of $H$ vanishes, zero energy is also the infinite-temperature canonical mean; the corresponding canonical mean of every $K_i$ is zero.

After removing the known one-dimensional target sector, central-spectrum adjacent-gap ratios rise from $0.542$ at $N=6$ to $0.596$ at $N=11$, close to the Gaussian unitary ensemble (GUE) benchmark.\cite{OganesyanHuse2007,Atas2013} The $Y$ terms generate complex coefficients, and no common antiunitary symmetry fixes all members of the independent term family. This numerical result supports interpreting the designed cluster eigenvector as an embedded scar at these sizes; it does not prove thermodynamic ETH or GUE universality. Stabilizer structure alone would not suffice, since exact thermal stabilizer eigenstates also exist.\cite{Hokkyo2026}

\subsection{Exact filtering identities without a commuting Hamiltonian}
Specify a photon-counting-type unraveling with jump operators
\begin{equation}
L_i=\sqrt{\gamma}\,Q_i,\qquad
\Heff=H-\frac{i\gamma}{2}D,\qquad D=\sum_i Q_i,\qquad \gamma>0.
\label{eq:jumps}
\end{equation}
Both the generator and its jumps must be specified: different detector unravelings of a Lindblad equation can give different no-click dynamics.\cite{Lindblad1976,Gorini1976,Dalibard1992,PlenioKnight1998,Minganti2019} In the graph basis $W\ket{s}$, $D$ counts syndrome defects. Thus
\begin{equation}
R\leq D\leq NR,\qquad HP=PH=DP=PD=0.
\label{eq:gap}
\end{equation}
There is no assumption that $[H,D]=0$.

Let $A(t)=\exp(-i\Heff t)=P+B(t)$, with $B=RAR$. For every $x\in R$, differentiation gives $\mathrm d\|B(t)x\|^2/\mathrm dt=-\gamma\bra{x}B^\dagger DB\ket{x}$. Equation~\eqref{eq:gap} then bounds the squared norm between $e^{-N\gamma t}\|x\|^2$ and $e^{-\gamma t}\|x\|^2$. This semigroup estimate controls the transient evolution directly, including when the effective Hamiltonian is nonnormal.

For any initial density matrix $\rho_0$, set $p=\tr(P\rho_0)$ and $\widetilde\rho(t)=A\rho_0A^\dagger$. The no-click probability $S$ and conditional fidelity satisfy the exact identities
\begin{equation}
S(t)=p+q(t),\qquad \Fid(t)=\frac{p}{S(t)},\qquad
(1-p)e^{-N\gamma t}\leq q(t)\leq(1-p)e^{-\gamma t}.
\label{eq:identity}
\end{equation}
In particular,
\begin{equation}
\Fid(t)\geq\frac{p}{p+(1-p)e^{-\gamma t}},\qquad
p\leq S(t)\leq p+(1-p)e^{-\gamma t}.
\label{eq:bound}
\end{equation}
For $p>0$, the accepted state converges to the cluster state with asymptotic probability $p$. If $p=0$, filtering cannot create the missing cluster component. For $0<p<1$ and $0<\epsilon_f<1$, a sufficient monitoring duration for fidelity at least $1-\epsilon_f$ is
\begin{equation}
t\geq\frac1\gamma\left[\log\frac{(1-p)(1-\epsilon_f)}{p\epsilon_f}\right]_+,
\label{eq:time}
\end{equation}
where $[x]_+=\max(0,x)$; for $p=1$, no filtering time is needed. A lower bound on $p$ suffices to use this formula. A no-click record by itself, without an input promise or a separate statistical verification protocol, does not give an unconditional fidelity certificate for arbitrary inputs.

\begin{figure}[htbp]
\centering\includegraphics[width=\linewidth]{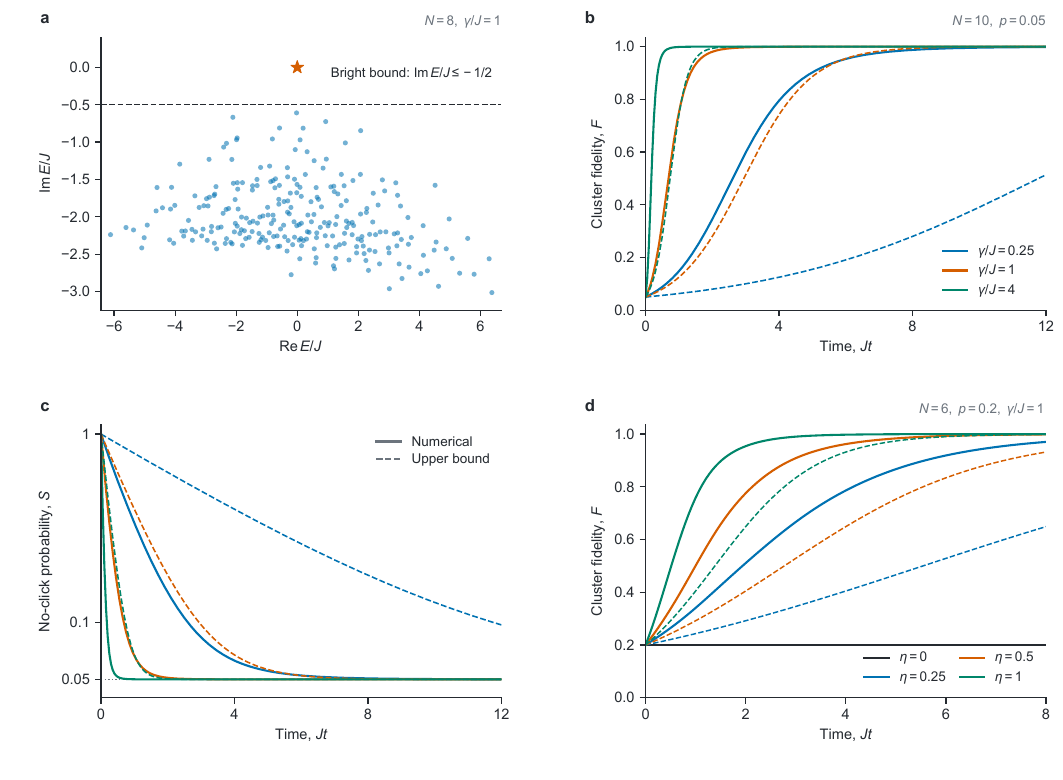}
\caption{\textbf{Conditional non-Hermitian filtering and detector inefficiency.} \textbf{a}, Full effective spectrum for $N=8$; the star is the exact dark eigenvalue and the dashed line bounds every bright imaginary part. \textbf{b,c}, Conditional fidelity and no-click probability for $N=10$ and initial overlap $p=0.05$. Solid curves are sparse matrix-exponential calculations; dashed curves are the fidelity lower bound and survival upper bound, respectively. The dotted line in \textbf{c} is $p$. \textbf{d}, Full conditional density-matrix evolution at $N=6$, $p=0.2$, $\gamma/J=1$, including unobserved jumps. Solid curves are numerical results and dashed curves use the guaranteed exponent $\eta\gamma$. $\eta=0$ is unconditional evolution and leaves the target weight unchanged.}
\label{fig:monitoring}
\end{figure}

All bright eigenvalues satisfy $\operatorname{Im}E\leq-\gamma/2$. For the $N=8$ example at $\gamma/J=1$, the largest bright imaginary part is $-0.60942J$, consistent with the universal bound. Direct evolution of the $N=10$ chain and the mixed-state detector calculation appear in Fig.~\ref{fig:monitoring}. Filtering faster than the universal bound reflects bright-sector dynamics and initial-state structure; the theorem does not require this faster observed rate.

\subsection{Which measurement is nondemolition?}
The binary target observable $P$ commutes with $H$ and every $L_i$. Both the dynamics and the measuring interaction therefore conserve it strictly in the ideal model, in the conventional nondemolition sense.\cite{Braginsky1980,Caves1980} The cluster state is itself dark and undisturbed. In general, $[H,K_i]\ne0$: the individual stabilizers need not be conserved on the complementary subspace while $H$ acts.

The unconditional Lindblad generator satisfies $\mathcal L^\dagger(P)=0$, so
\begin{equation}
\tr[P\rho(t)]=p.
\label{eq:unconditional}
\end{equation}
The generator is unital, and both $P$ and $R/(2^N-1)$ are stationary. It therefore does not autonomously prepare the target from arbitrary inputs. Its one-dimensional dark sector is a special case of decoherence-free structure,\cite{ZanardiRasetti1997,Lidar1998,TicozziViola2008,AlbertJiang2014} but the measurement does not pump population into that sector as engineered dissipators do.\cite{Kraus2008,Diehl2008,Verstraete2009,TicozziViola2012,Wang2024DFS}

A coarse stabilizer-parity measurement mediated by an ancilla provides a suitable microscopic realization. Its projectors $(\id\pm K_i)/2$ preserve $\Cstate$, whereas resolving all constituent one-qubit outcomes generally destroys it. Established nondemolition verification schemes use this distinction.\cite{Liu2021QND} Supplementary Notes~1 and 5 give the no-go argument for unrestricted single-product-basis certification and an explicit weak-measurement instrument realizing Eq.~\eqref{eq:jumps}.

\subsection{Inefficiency, coherent errors and resource constraints}
At detector efficiency $\eta$, conditioning on no observed clicks requires keeping the unobserved jumps in the equation:
\begin{equation}
\dot{\widetilde\rho}=-i[H,\widetilde\rho]+(1-\eta)\gamma\sum_iQ_i\widetilde\rho Q_i
-\frac\gamma2\{D,\widetilde\rho\}.
\label{eq:eta}
\end{equation}
The target weight remains $p$, and the bright trace obeys $\dot q=-\eta\gamma\tr(D\widetilde\rho_R)$. The fidelity and survival bounds therefore hold with the slow rate $\gamma$ replaced by $\eta\gamma$ and the fast rate $N\gamma$ replaced by $N\eta\gamma$. Replacing $\gamma\mapsto\eta\gamma$ in a pure-state effective Hamiltonian alone does not give the correct physical model for missed clicks.\cite{Brun2002,JacobsSteck2006}

\begin{figure}[htbp]
\centering\includegraphics[width=\linewidth]{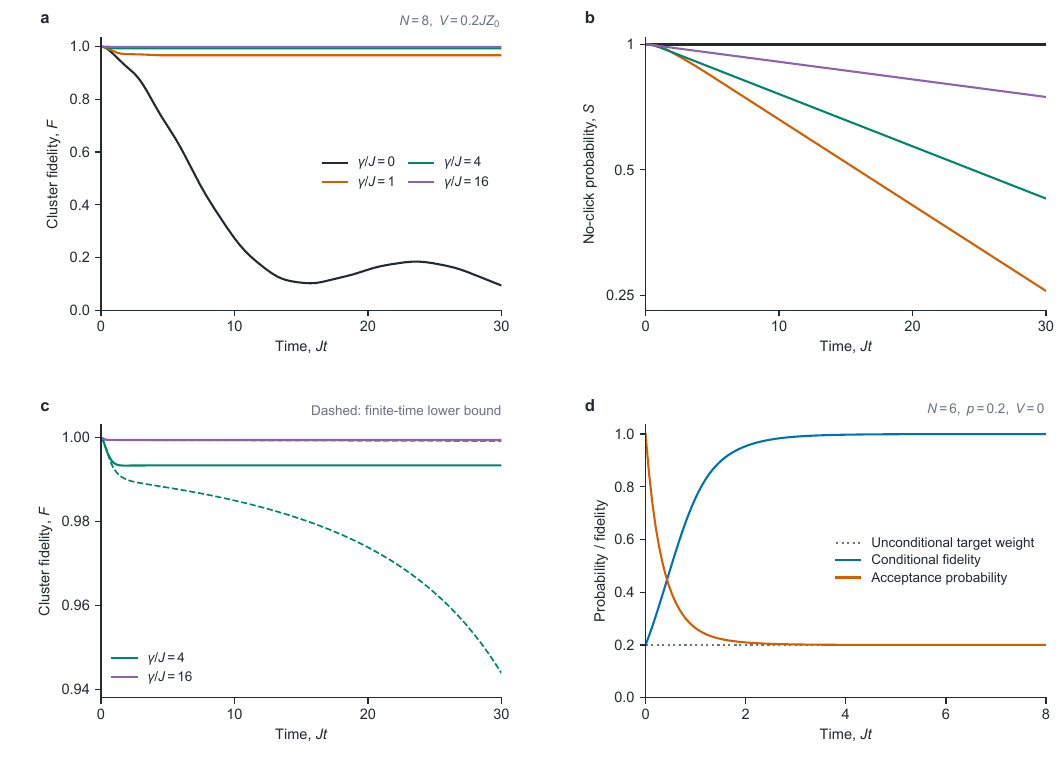}
\caption{\textbf{Limits of measurement protection.} \textbf{a,b}, Conditional fidelity and survival for an initially exact $N=8$ cluster state under the physical error $V=0.2JZ_0$, for the stated monitoring rates. \textbf{c}, Numerical fidelities (solid) and the conservative finite-time bounds in Eq.~\eqref{eq:errorbound} (dashed), shown on an expanded fidelity scale from 0.94 to 1.00. Once $a_*$ vanishes, the guarantee becomes trivial. \textbf{d}, An unperturbed $N=6$, $p=0.2$ example contrasts the unchanged unconditional target population with the increasing conditional fidelity and decreasing acceptance probability at $\gamma/J=1$. All panels use actual matrix-exponential evolution, not fitted curves.}
\label{fig:errors}
\end{figure}

Protection against coherent Hamiltonian errors has a separate limitation. Let $V=V^\dagger$, start from the exact target with perfect detection, and define $v=\|RVP\|$ and $\alpha=\gamma/2$. With
\begin{equation}
 b_*(t)=\frac v\alpha(1-e^{-\alpha t}),\qquad
 a_*(t)=\left[1-\frac{v^2}{\alpha}\left(t-\frac{1-e^{-\alpha t}}\alpha\right)\right]_+,
\label{eq:error}
\end{equation}
Duhamel's formula and contractivity imply
\begin{equation}
\Fid(t)\geq\frac{a_*^2}{a_*^2+b_*^2},\qquad S(t)\geq a_*^2
\label{eq:errorbound}
\end{equation}
when $a_*>0$; otherwise the lower bound is only trivial. Diagonal blocks of $V$ can be included in the Hermitian sector Hamiltonians. This bound holds at finite times and does not provide indefinite error correction. The numerical example $V=0.2JZ_0$ shows suppressed conditional leakage with nonunit survival (Fig.~\ref{fig:errors}), consistent with the connection between strong monitoring and Zeno dynamics.\cite{MisraSudarshan1977,FacchiPascazio2002,FacchiPascazio2008}

The rate budget sets the guarantee. For nonuniform rates $\gamma_i$, the defect operator $\sum_i\gamma_iQ_i$ has bright gap $g=\min_i\gamma_i$. At fixed $\Gamma=\sum_i\gamma_i$, $g\leq\Gamma/N$, so uniform allocation maximizes the Hamiltonian-independent guarantee. It need not optimize the actual decay for a fixed noncommuting Hamiltonian. The guarantee is tight over the broader class with $H=0$ and a single-defect input. A size-independent exponent at fixed per-check rate requires an extensive total rate $N\gamma$.

\begin{figure}[htbp]
\centering\includegraphics[width=\linewidth]{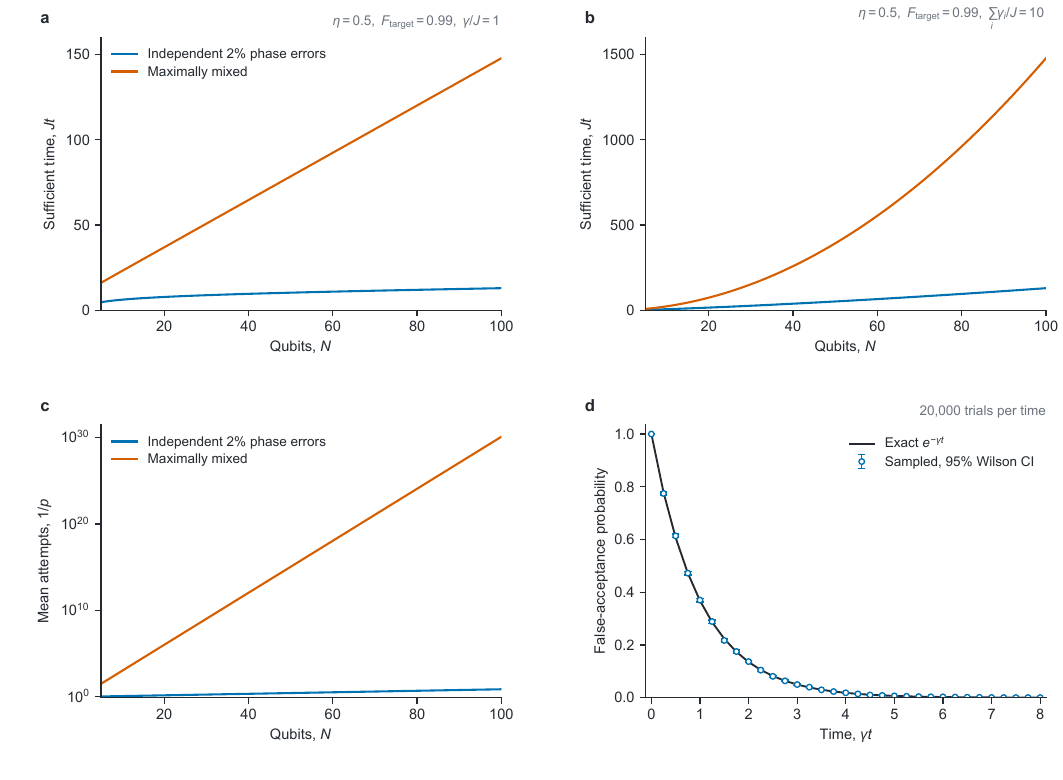}
\caption{\textbf{Monitoring duration and acceptance costs.} \textbf{a,b}, Sufficient times for fidelity $0.99$ at efficiency $\eta=0.5$, evaluated from Eq.~\eqref{eq:time}, with fixed per-check rate and fixed total rate, respectively. These are analytic sufficient-time curves. \textbf{c}, Asymptotic independent-attempt cost $1/p$ for the same two initial-state families. \textbf{d}, Monte Carlo Bernoulli sampling of the tight worst-case false-acceptance law $e^{-\gamma t}$, realized by a single defect with $H=0$. These samples illustrate finite-count fluctuations and do not independently validate the dynamics. There are 20,000 independent trials per plotted time; error bars are pointwise nominal 95\% Wilson intervals, not simultaneous confidence bands or hardware data.}
\label{fig:resources}
\end{figure}

Figure~\ref{fig:resources} plots the exact resource formulas up to $N=100$, without simulating 100-qubit interacting dynamics. Independent phase errors with probability $r$ give $p=(1-r)^N$, while a maximally mixed input gives $p=2^{-N}$. The asymptotic number of independent attempts per retained target is $1/p$. At finite time, the requirement $\Fid\geq1-\epsilon_f$ and identity $\Fid S=p$ give $S\leq p/(1-\epsilon_f)$, so the preparation cost remains exponential. For $N=100$, the two example asymptotic costs are approximately $7.54$ attempts for $r=0.02$ and $1.27\times10^{30}$ for a maximally mixed input.

\FloatBarrier
\section{Discussion}
The construction gives a target-preserving filter for a local embedded cluster eigenstate with an explicitly defined measurement process. Its accuracy follows from a positive operator gap and invariant target weight; neither a commuting bright Hamiltonian nor a diagonalizable effective generator is needed. The argument also bounds false acceptance for an orthogonal input: the probability of a false no-click acceptance is at most $e^{-\eta\gamma t}$. High output fidelity is not guaranteed for a state with zero target component. Statistical verification of an untrusted source further requires a sampling model, as in established verification theory.\cite{Pallister2018,ZhuHayashi2019Adversarial,Yu2022}

The construction has several limits. A dark cluster eigenstate alone does not establish a scar tower, revival dynamics, a many-body phase transition, topological protection or an experimental implementation. The finite-size spectral and entanglement diagnostics support the scar interpretation, while thermodynamic thermalization remains unresolved. Non-Hermitian filtering is an established mechanism, with direct antecedents in decoherence-free scar embeddings, non-Hermitian scar stabilization and nondemolition stabilizer verification.\cite{Wang2024DFS,Omiya2025,Liu2021QND} The rate guarantee also holds for $H=0$, so it does not establish an advantage caused by chaos. The present benchmark gives an explicit example that brings together local many-body evolution, an exact target, reproducible finite-size evidence, and explicit filtering guarantees and costs.

For a higher-dimensional protected subspace, the detector must act identically on all encoded states to preserve arbitrary superpositions. Preserving each basis vector alone is insufficient. Physical detector errors must also be specified through their jump operators; efficiency loss and coherent calibration errors do not encompass arbitrary noise. Any improvement in verification efficiency over existing graph-state protocols would need to be established through a matched resource comparison that accounts for ancillas, local interactions, repetitions and the prior target overlap. We do not claim a universal improvement.

\section{Methods}
\subsection{Numerical conventions and reproducibility}
We set $\hbar=J=1$. Bit $i$ is the $i$th least significant binary digit. The coefficient array has shape $(N,2,3)$, with direction order $(-1,+1)$ and Pauli order $(X,Y,Z)$. NumPy's default generator uses seed $91000+100N+s$ for realization $s$, starting at zero. The pure-trajectory and detector-efficiency examples use state seeds 404 and 405, respectively. The complementary random component is a normalized complex Gaussian vector, and the target amplitude $\sqrt p$ is fixed separately. Parameters were not selected by searching for favourable spectral statistics.

We obtain Hermitian spectra by dense diagonalization of the exact $(2^N-1)$-dimensional complementary block. Adjacent-gap statistics use the central 50\% by eigenvalue index, with averages taken first within each realization. The displayed intervals come from four thousand bootstrap resamples of the realization means. Entropy is computed by singular-value decomposition across the half-ring cut after the full Clifford transformation into the physical frame. We use natural logarithms throughout. The small number of realizations and the changing half-system dimension for odd $N$ do not support an asymptotic fit. Page entropy serves as a background reference, rather than an assumed exact value for the finite-energy interacting system.\cite{Page1993}

Non-Hermitian state evolution and conditional density matrices are computed with sparse actions of the matrix exponential.\cite{AlMohyHigham2011} The full density calculation uses column-major vectorization and includes missed jumps explicitly. The main dynamical curves use neither a Trotter approximation nor tensor-network truncation. Supplementary Fig.~S1 separately tests convergence of an exact finite-step Kraus instrument to the continuous equation. Independent checks use a Kronecker construction, direct density-matrix integration and a pure-state/density-matrix comparison. The software stack comprises NumPy, SciPy and Matplotlib.\cite{Harris2020,Virtanen2020,Hunter2007} The source package includes exact versions, seeds, arrays, CSV data and test reports.

\subsection{Computer-checked matrix dynamics}
The accompanying Lean 4 project uses fixed versions Lean 4.19.0 and mathlib v4.19.0.\cite{Lean2021,Mathlib2020} It constructs the ring CZ/Hadamard circuit and physical Pauli Hamiltonian, proves the dark-projector and loss-gap identities, and derives filtering bounds for the actual conditional matrix exponential with arbitrary positive semidefinite, unit-trace inputs, including target--bright coherences. It proves that the missed-jump dynamics is completely positive and trace-nonincreasing in the physical parameter range. The project also connects the sequential finite measurement instrument to its continuous limit, proves arbitrary-cut entanglement witnesses, and applies the coherent-error bound to the actual ring Hamiltonian. The coverage map and compiler/axiom audits specify the statements proved and their scope. These proofs establish mathematical claims under the stated assumptions; floating-point output, many-body thermalization and scientific novelty require separate evidence.

\subsection{Scope of the evidence}
Supplementary Notes~1--8 derive the analytic statements and state their assumptions. Supplementary Notes~9--11 describe the numerical procedures, measurement-window compatibility, model limitations and comparisons with prior work. All plotted interacting dynamics are for $N\leq10$; the largest Hermitian diagonalization uses $N=11$. The study reports no experimental observations, hardware access, trained classification model or thermodynamic numerical simulation.

\section{Data availability}
The accompanying source package provides all newly generated data used in the figures in machine-readable CSV, NPY and NPZ formats, together with parameter and verification records. It contains no redistributed third-party article PDFs.
\section{Code availability}
The accompanying package contains the model, complete simulation and plotting scripts, independent verification routines, environment specification, build instructions and both LaTeX manuscripts, together with the Lean source project and its compiler and axiom-audit records. Once the documented dependencies are installed, the physics calculations can be reproduced without network access.

\FloatBarrier
\bibliography{references}

\begin{thebibliography}{10}
\expandafter\ifx\csname url\endcsname\relax
  \def\url#1{\texttt{#1}}\fi
\expandafter\ifx\csname urlprefix\endcsname\relax\def\urlprefix{URL }\fi
\providecommand{\bibinfo}[2]{#2}
\providecommand{\eprint}[2][]{\url{#2}}

\bibitem{Deutsch1991}
\bibinfo{author}{Deutsch, J.~M.}
\newblock \bibinfo{title}{{Quantum statistical mechanics in a closed system}}.
\newblock \emph{\bibinfo{journal}{Physical Review A}}
  \textbf{\bibinfo{volume}{43}}, \bibinfo{pages}{2046--2049}
  (\bibinfo{year}{1991}).
\newblock \urlprefix\url{https://doi.org/10.1103/physreva.43.2046}.

\bibitem{Srednicki1994}
\bibinfo{author}{Srednicki, M.}
\newblock \bibinfo{title}{{Chaos and quantum thermalization}}.
\newblock \emph{\bibinfo{journal}{Physical Review E}}
  \textbf{\bibinfo{volume}{50}}, \bibinfo{pages}{888--901}
  (\bibinfo{year}{1994}).
\newblock \urlprefix\url{https://doi.org/10.1103/physreve.50.888}.

\bibitem{Rigol2008}
\bibinfo{author}{Rigol, M.}, \bibinfo{author}{Dunjko, V.} \&
  \bibinfo{author}{Olshanii, M.}
\newblock \bibinfo{title}{{Thermalization and its mechanism for generic
  isolated quantum systems}}.
\newblock \emph{\bibinfo{journal}{Nature}} \textbf{\bibinfo{volume}{452}},
  \bibinfo{pages}{854--858} (\bibinfo{year}{2008}).
\newblock \urlprefix\url{https://doi.org/10.1038/nature06838}.

\bibitem{DAlessio2016}
\bibinfo{author}{D'Alessio, L.}, \bibinfo{author}{Kafri, Y.},
  \bibinfo{author}{Polkovnikov, A.} \& \bibinfo{author}{Rigol, M.}
\newblock \bibinfo{title}{{From quantum chaos and eigenstate thermalization to
  statistical mechanics and thermodynamics}}.
\newblock \emph{\bibinfo{journal}{Advances in Physics}}
  \textbf{\bibinfo{volume}{65}}, \bibinfo{pages}{239--362}
  (\bibinfo{year}{2016}).
\newblock \urlprefix\url{https://doi.org/10.1080/00018732.2016.1198134}.

\bibitem{Mori2018}
\bibinfo{author}{Mori, T.}, \bibinfo{author}{Ikeda, T.~N.},
  \bibinfo{author}{Kaminishi, E.} \& \bibinfo{author}{Ueda, M.}
\newblock \bibinfo{title}{{Thermalization and prethermalization in isolated
  quantum systems: a theoretical overview}}.
\newblock \emph{\bibinfo{journal}{Journal of Physics B: Atomic, Molecular and
  Optical Physics}} \textbf{\bibinfo{volume}{51}}, \bibinfo{pages}{112001}
  (\bibinfo{year}{2018}).
\newblock \urlprefix\url{https://doi.org/10.1088/1361-6455/aabcdf}.

\bibitem{Bernien2017}
\bibinfo{author}{Bernien, H.} \emph{et~al.}
\newblock \bibinfo{title}{{Probing many-body dynamics on a 51-atom quantum
  simulator}}.
\newblock \emph{\bibinfo{journal}{Nature}} \textbf{\bibinfo{volume}{551}},
  \bibinfo{pages}{579--584} (\bibinfo{year}{2017}).
\newblock \urlprefix\url{https://doi.org/10.1038/nature24622}.

\bibitem{Turner2018}
\bibinfo{author}{Turner, C.~J.}, \bibinfo{author}{Michailidis, A.~A.},
  \bibinfo{author}{Abanin, D.~A.}, \bibinfo{author}{Serbyn, M.} \&
  \bibinfo{author}{Papi{\'{c}}, Z.}
\newblock \bibinfo{title}{{Weak ergodicity breaking from quantum many-body
  scars}}.
\newblock \emph{\bibinfo{journal}{Nature Physics}}
  \textbf{\bibinfo{volume}{14}}, \bibinfo{pages}{745--749}
  (\bibinfo{year}{2018}).
\newblock \urlprefix\url{https://doi.org/10.1038/s41567-018-0137-5}.

\bibitem{Turner2018PRB}
\bibinfo{author}{Turner, C.~J.}, \bibinfo{author}{Michailidis, A.~A.},
  \bibinfo{author}{Abanin, D.~A.}, \bibinfo{author}{Serbyn, M.} \&
  \bibinfo{author}{Papi{\'{c}}, Z.}
\newblock \bibinfo{title}{{Quantum scarred eigenstates in a Rydberg atom chain:
  Entanglement, breakdown of thermalization, and stability to perturbations}}.
\newblock \emph{\bibinfo{journal}{Physical Review B}}
  \textbf{\bibinfo{volume}{98}}, \bibinfo{pages}{155134}
  (\bibinfo{year}{2018}).
\newblock \urlprefix\url{https://doi.org/10.1103/physrevb.98.155134}.

\bibitem{Serbyn2021}
\bibinfo{author}{Serbyn, M.}, \bibinfo{author}{Abanin, D.~A.} \&
  \bibinfo{author}{Papi{\'{c}}, Z.}
\newblock \bibinfo{title}{{Quantum many-body scars and weak breaking of
  ergodicity}}.
\newblock \emph{\bibinfo{journal}{Nature Physics}}
  \textbf{\bibinfo{volume}{17}}, \bibinfo{pages}{675--685}
  (\bibinfo{year}{2021}).
\newblock \urlprefix\url{https://doi.org/10.1038/s41567-021-01230-2}.

\bibitem{Moudgalya2022}
\bibinfo{author}{Moudgalya, S.}, \bibinfo{author}{Bernevig, B.~A.} \&
  \bibinfo{author}{Regnault, N.}
\newblock \bibinfo{title}{{Quantum many-body scars and Hilbert space
  fragmentation: a review of exact results}}.
\newblock \emph{\bibinfo{journal}{Reports on Progress in Physics}}
  \textbf{\bibinfo{volume}{85}}, \bibinfo{pages}{086501}
  (\bibinfo{year}{2022}).
\newblock \urlprefix\url{https://doi.org/10.1088/1361-6633/ac73a0}.

\bibitem{Chandran2023}
\bibinfo{author}{Chandran, A.}, \bibinfo{author}{Iadecola, T.},
  \bibinfo{author}{Khemani, V.} \& \bibinfo{author}{Moessner, R.}
\newblock \bibinfo{title}{{Quantum Many-Body Scars: A Quasiparticle
  Perspective}}.
\newblock \emph{\bibinfo{journal}{Annual Review of Condensed Matter Physics}}
  \textbf{\bibinfo{volume}{14}}, \bibinfo{pages}{443--469}
  (\bibinfo{year}{2023}).
\newblock
  \urlprefix\url{https://doi.org/10.1146/annurev-conmatphys-031620-101617}.

\bibitem{ShiraishiMori2017}
\bibinfo{author}{Shiraishi, N.} \& \bibinfo{author}{Mori, T.}
\newblock \bibinfo{title}{{Systematic Construction of Counterexamples to the
  Eigenstate Thermalization Hypothesis}}.
\newblock \emph{\bibinfo{journal}{Physical Review Letters}}
  \textbf{\bibinfo{volume}{119}}, \bibinfo{pages}{030601}
  (\bibinfo{year}{2017}).
\newblock \urlprefix\url{https://doi.org/10.1103/physrevlett.119.030601}.

\bibitem{LinMotrunich2019}
\bibinfo{author}{Lin, C.-J.} \& \bibinfo{author}{Motrunich, O.~I.}
\newblock \bibinfo{title}{{Exact Quantum Many-Body Scar States in the
  Rydberg-Blockaded Atom Chain}}.
\newblock \emph{\bibinfo{journal}{Physical Review Letters}}
  \textbf{\bibinfo{volume}{122}}, \bibinfo{pages}{173401}
  (\bibinfo{year}{2019}).
\newblock \urlprefix\url{https://doi.org/10.1103/physrevlett.122.173401}.

\bibitem{Moudgalya2018}
\bibinfo{author}{Moudgalya, S.}, \bibinfo{author}{Rachel, S.},
  \bibinfo{author}{Bernevig, B.~A.} \& \bibinfo{author}{Regnault, N.}
\newblock \bibinfo{title}{{Exact excited states of nonintegrable models}}.
\newblock \emph{\bibinfo{journal}{Physical Review B}}
  \textbf{\bibinfo{volume}{98}}, \bibinfo{pages}{235155}
  (\bibinfo{year}{2018}).
\newblock \urlprefix\url{https://doi.org/10.1103/physrevb.98.235155}.

\bibitem{SchecterIadecola2019}
\bibinfo{author}{Schecter, M.} \& \bibinfo{author}{Iadecola, T.}
\newblock \bibinfo{title}{{Weak Ergodicity Breaking and Quantum Many-Body Scars
  in Spin-1 XY Magnets}}.
\newblock \emph{\bibinfo{journal}{Physical Review Letters}}
  \textbf{\bibinfo{volume}{123}}, \bibinfo{pages}{147201}
  (\bibinfo{year}{2019}).
\newblock \urlprefix\url{https://doi.org/10.1103/physrevlett.123.147201}.

\bibitem{Mark2020}
\bibinfo{author}{Mark, D.~K.}, \bibinfo{author}{Lin, C.-J.} \&
  \bibinfo{author}{Motrunich, O.~I.}
\newblock \bibinfo{title}{{Unified structure for exact towers of scar states in
  the Affleck-Kennedy-Lieb-Tasaki and other models}}.
\newblock \emph{\bibinfo{journal}{Physical Review B}}
  \textbf{\bibinfo{volume}{101}}, \bibinfo{pages}{195131}
  (\bibinfo{year}{2020}).
\newblock \urlprefix\url{https://doi.org/10.1103/physrevb.101.195131}.

\bibitem{BriegelRaussendorf2001}
\bibinfo{author}{Briegel, H.~J.} \& \bibinfo{author}{Raussendorf, R.}
\newblock \bibinfo{title}{{Persistent Entanglement in Arrays of Interacting
  Particles}}.
\newblock \emph{\bibinfo{journal}{Physical Review Letters}}
  \textbf{\bibinfo{volume}{86}}, \bibinfo{pages}{910--913}
  (\bibinfo{year}{2001}).
\newblock \urlprefix\url{https://doi.org/10.1103/physrevlett.86.910}.

\bibitem{RaussendorfBriegel2001}
\bibinfo{author}{Raussendorf, R.} \& \bibinfo{author}{Briegel, H.~J.}
\newblock \bibinfo{title}{{A One-Way Quantum Computer}}.
\newblock \emph{\bibinfo{journal}{Physical Review Letters}}
  \textbf{\bibinfo{volume}{86}}, \bibinfo{pages}{5188--5191}
  (\bibinfo{year}{2001}).
\newblock \urlprefix\url{https://doi.org/10.1103/physrevlett.86.5188}.

\bibitem{Raussendorf2003}
\bibinfo{author}{Raussendorf, R.}, \bibinfo{author}{Browne, D.~E.} \&
  \bibinfo{author}{Briegel, H.~J.}
\newblock \bibinfo{title}{{Measurement-based quantum computation on cluster
  states}}.
\newblock \emph{\bibinfo{journal}{Physical Review A}}
  \textbf{\bibinfo{volume}{68}}, \bibinfo{pages}{022312}
  (\bibinfo{year}{2003}).
\newblock \urlprefix\url{https://doi.org/10.1103/physreva.68.022312}.

\bibitem{Hein2004}
\bibinfo{author}{Hein, M.}, \bibinfo{author}{Eisert, J.} \&
  \bibinfo{author}{Briegel, H.~J.}
\newblock \bibinfo{title}{{Multiparty entanglement in graph states}}.
\newblock \emph{\bibinfo{journal}{Physical Review A}}
  \textbf{\bibinfo{volume}{69}}, \bibinfo{pages}{062311}
  (\bibinfo{year}{2004}).
\newblock \urlprefix\url{https://doi.org/10.1103/physreva.69.062311}.

\bibitem{StabilizerScars2025}
\bibinfo{author}{Hartse, J.}, \bibinfo{author}{Fidkowski, L.} \&
  \bibinfo{author}{Mueller, N.}
\newblock \bibinfo{title}{{Stabilizer Scars}}.
\newblock \emph{\bibinfo{journal}{Physical Review Letters}}
  \textbf{\bibinfo{volume}{135}}, \bibinfo{pages}{060402}
  (\bibinfo{year}{2025}).
\newblock \urlprefix\url{https://doi.org/10.1103/n5hb-l5p5}.

\bibitem{Dooley2026}
\bibinfo{author}{Dooley, S.}
\newblock \bibinfo{title}{{Parent Hamiltonians for Stabilizer Quantum Many-Body
  Scars}}.
\newblock \emph{\bibinfo{journal}{Physical Review Letters}}
  \textbf{\bibinfo{volume}{136}}, \bibinfo{pages}{240402}
  (\bibinfo{year}{2026}).
\newblock \urlprefix\url{https://doi.org/10.1103/3vk5-483c}.

\bibitem{Larsen2026}
\bibinfo{author}{Larsen, P.~G.}, \bibinfo{author}{Nielsen, A. E.~B.},
  \bibinfo{author}{Eckardt, A.} \& \bibinfo{author}{Petiziol, F.}
\newblock \bibinfo{title}{{Experimental protocol for observing single quantum
  many-body scars with transmon qubits}}.
\newblock \emph{\bibinfo{journal}{SciPost Physics}}
  \textbf{\bibinfo{volume}{20}}, \bibinfo{pages}{036} (\bibinfo{year}{2026}).
\newblock \urlprefix\url{https://doi.org/10.21468/scipostphys.20.2.036}.

\bibitem{Hartse2026Benchmark}
\bibinfo{author}{Hartse, J.}, \bibinfo{author}{Raza, M.},
  \bibinfo{author}{Shravan, S.}, \bibinfo{author}{Deutsch, I.~H.} \&
  \bibinfo{author}{Mueller, N.}
\newblock \bibinfo{title}{{Benchmarking quantum simulation at scale}}
  (\bibinfo{year}{2026}).
\newblock \urlprefix\url{https://arxiv.org/abs/2607.14212v1}.
\newblock \bibinfo{note}{Preprint}, \eprint{2607.14212v1}.

\bibitem{Li2026Fingerprint}
\bibinfo{author}{Li, R.} \emph{et~al.}
\newblock \bibinfo{title}{{Large Scale Entanglement Structure Detection in
  100-Qubit Systems via Local Joint Measurements}} (\bibinfo{year}{2026}).
\newblock \urlprefix\url{https://arxiv.org/abs/2608.20170v1}.
\newblock \bibinfo{note}{Preprint}, \eprint{2608.20170v1}.

\bibitem{Liu2021QND}
\bibinfo{author}{Liu, Y.-C.}, \bibinfo{author}{Shang, J.},
  \bibinfo{author}{Han, R.} \& \bibinfo{author}{Zhang, X.}
\newblock \bibinfo{title}{{Universally Optimal Verification of Entangled States
  with Nondemolition Measurements}}.
\newblock \emph{\bibinfo{journal}{Physical Review Letters}}
  \textbf{\bibinfo{volume}{126}}, \bibinfo{pages}{090504}
  (\bibinfo{year}{2021}).
\newblock \urlprefix\url{https://doi.org/10.1103/physrevlett.126.090504}.

\bibitem{Dangniam2020}
\bibinfo{author}{Dangniam, N.}, \bibinfo{author}{Han, Y.-G.} \&
  \bibinfo{author}{Zhu, H.}
\newblock \bibinfo{title}{{Optimal verification of stabilizer states}}.
\newblock \emph{\bibinfo{journal}{Physical Review Research}}
  \textbf{\bibinfo{volume}{2}}, \bibinfo{pages}{043323} (\bibinfo{year}{2020}).
\newblock \urlprefix\url{https://doi.org/10.1103/physrevresearch.2.043323}.

\bibitem{Riera2023}
\bibinfo{author}{Riera-S{\`{a}}bat, F.}, \bibinfo{author}{Miguel-Ramiro, J.} \&
  \bibinfo{author}{D{\"{u}}r, W.}
\newblock \bibinfo{title}{{Nondestructive verification of entangled states via
  fidelity witnessing}}.
\newblock \emph{\bibinfo{journal}{Physical Review A}}
  \textbf{\bibinfo{volume}{107}}, \bibinfo{pages}{022414}
  (\bibinfo{year}{2023}).
\newblock \urlprefix\url{https://doi.org/10.1103/physreva.107.022414}.

\bibitem{Wang2024DFS}
\bibinfo{author}{Wang, H.-R.} \emph{et~al.}
\newblock \bibinfo{title}{{Embedding Quantum Many-Body Scars into
  Decoherence-Free Subspaces}}.
\newblock \emph{\bibinfo{journal}{Physical Review Letters}}
  \textbf{\bibinfo{volume}{132}}, \bibinfo{pages}{150401}
  (\bibinfo{year}{2024}).
\newblock \urlprefix\url{https://doi.org/10.1103/physrevlett.132.150401}.

\bibitem{Chen2023NH}
\bibinfo{author}{Chen, Q.}, \bibinfo{author}{Chen, S.~A.} \&
  \bibinfo{author}{Zhu, Z.}
\newblock \bibinfo{title}{{Weak ergodicity breaking in non-Hermitian many-body
  systems}}.
\newblock \emph{\bibinfo{journal}{SciPost Physics}}
  \textbf{\bibinfo{volume}{15}}, \bibinfo{pages}{052} (\bibinfo{year}{2023}).
\newblock \urlprefix\url{https://doi.org/10.21468/scipostphys.15.2.052}.

\bibitem{Omiya2025}
\bibinfo{author}{Omiya, K.} \& \bibinfo{author}{Nakagawa, Y.~O.}
\newblock \bibinfo{title}{{Non-Hermitian Quantum Many-Body Scar Phase}}
  (\bibinfo{year}{2025}).
\newblock \urlprefix\url{https://arxiv.org/abs/2507.22583v1}.
\newblock \bibinfo{note}{Preprint}, \eprint{2507.22583v1}.

\bibitem{Moudgalya2020MPS}
\bibinfo{author}{Moudgalya, S.}, \bibinfo{author}{O'Brien, E.},
  \bibinfo{author}{Bernevig, B.~A.}, \bibinfo{author}{Fendley, P.} \&
  \bibinfo{author}{Regnault, N.}
\newblock \bibinfo{title}{{Large classes of quantum scarred Hamiltonians from
  matrix product states}}.
\newblock \emph{\bibinfo{journal}{Physical Review B}}
  \textbf{\bibinfo{volume}{102}}, \bibinfo{pages}{085120}
  (\bibinfo{year}{2020}).
\newblock \urlprefix\url{https://doi.org/10.1103/physrevb.102.085120}.

\bibitem{OganesyanHuse2007}
\bibinfo{author}{Oganesyan, V.} \& \bibinfo{author}{Huse, D.~A.}
\newblock \bibinfo{title}{{Localization of interacting fermions at high
  temperature}}.
\newblock \emph{\bibinfo{journal}{Physical Review B}}
  \textbf{\bibinfo{volume}{75}}, \bibinfo{pages}{155111}
  (\bibinfo{year}{2007}).
\newblock \urlprefix\url{https://doi.org/10.1103/physrevb.75.155111}.

\bibitem{Atas2013}
\bibinfo{author}{Atas, Y.~Y.}, \bibinfo{author}{Bogomolny, E.},
  \bibinfo{author}{Giraud, O.} \& \bibinfo{author}{Roux, G.}
\newblock \bibinfo{title}{{Distribution of the Ratio of Consecutive Level
  Spacings in Random Matrix Ensembles}}.
\newblock \emph{\bibinfo{journal}{Physical Review Letters}}
  \textbf{\bibinfo{volume}{110}}, \bibinfo{pages}{084101}
  (\bibinfo{year}{2013}).
\newblock \urlprefix\url{https://doi.org/10.1103/physrevlett.110.084101}.

\bibitem{Hokkyo2026}
\bibinfo{author}{Hokkyo, A.}
\newblock \bibinfo{title}{{Exact Thermal Stabilizer Eigenstates at Infinite
  Temperature}} (\bibinfo{year}{2026}).
\newblock \urlprefix\url{https://arxiv.org/abs/2601.16177v2}.
\newblock \bibinfo{note}{Preprint}, \eprint{2601.16177v2}.

\bibitem{Lindblad1976}
\bibinfo{author}{Lindblad, G.}
\newblock \bibinfo{title}{{On the generators of quantum dynamical semigroups}}.
\newblock \emph{\bibinfo{journal}{Communications in Mathematical Physics}}
  \textbf{\bibinfo{volume}{48}}, \bibinfo{pages}{119--130}
  (\bibinfo{year}{1976}).
\newblock \urlprefix\url{https://doi.org/10.1007/bf01608499}.

\bibitem{Gorini1976}
\bibinfo{author}{Gorini, V.}, \bibinfo{author}{Kossakowski, A.} \&
  \bibinfo{author}{Sudarshan, E. C.~G.}
\newblock \bibinfo{title}{{Completely positive dynamical semigroups of N-level
  systems}}.
\newblock \emph{\bibinfo{journal}{Journal of Mathematical Physics}}
  \textbf{\bibinfo{volume}{17}}, \bibinfo{pages}{821--825}
  (\bibinfo{year}{1976}).
\newblock \urlprefix\url{https://doi.org/10.1063/1.522979}.

\bibitem{Dalibard1992}
\bibinfo{author}{Dalibard, J.}, \bibinfo{author}{Castin, Y.} \&
  \bibinfo{author}{M{\o}lmer, K.}
\newblock \bibinfo{title}{{Wave-function approach to dissipative processes in
  quantum optics}}.
\newblock \emph{\bibinfo{journal}{Physical Review Letters}}
  \textbf{\bibinfo{volume}{68}}, \bibinfo{pages}{580--583}
  (\bibinfo{year}{1992}).
\newblock \urlprefix\url{https://doi.org/10.1103/physrevlett.68.580}.

\bibitem{PlenioKnight1998}
\bibinfo{author}{Plenio, M.~B.} \& \bibinfo{author}{Knight, P.~L.}
\newblock \bibinfo{title}{{The quantum-jump approach to dissipative dynamics in
  quantum optics}}.
\newblock \emph{\bibinfo{journal}{Reviews of Modern Physics}}
  \textbf{\bibinfo{volume}{70}}, \bibinfo{pages}{101--144}
  (\bibinfo{year}{1998}).
\newblock \urlprefix\url{https://doi.org/10.1103/revmodphys.70.101}.

\bibitem{Minganti2019}
\bibinfo{author}{Minganti, F.}, \bibinfo{author}{Miranowicz, A.},
  \bibinfo{author}{Chhajlany, R.~W.} \& \bibinfo{author}{Nori, F.}
\newblock \bibinfo{title}{{Quantum exceptional points of non-Hermitian
  Hamiltonians and Liouvillians: The effects of quantum jumps}}.
\newblock \emph{\bibinfo{journal}{Physical Review A}}
  \textbf{\bibinfo{volume}{100}}, \bibinfo{pages}{062131}
  (\bibinfo{year}{2019}).
\newblock \urlprefix\url{https://doi.org/10.1103/physreva.100.062131}.

\bibitem{Braginsky1980}
\bibinfo{author}{Braginsky, V.~B.}, \bibinfo{author}{Vorontsov, Y.~I.} \&
  \bibinfo{author}{Thorne, K.~S.}
\newblock \bibinfo{title}{{Quantum Nondemolition Measurements}}.
\newblock \emph{\bibinfo{journal}{Science}} \textbf{\bibinfo{volume}{209}},
  \bibinfo{pages}{547--557} (\bibinfo{year}{1980}).
\newblock \urlprefix\url{https://doi.org/10.1126/science.209.4456.547}.

\bibitem{Caves1980}
\bibinfo{author}{Caves, C.~M.}, \bibinfo{author}{Thorne, K.~S.},
  \bibinfo{author}{Drever, R. W.~P.}, \bibinfo{author}{Sandberg, V.~D.} \&
  \bibinfo{author}{Zimmermann, M.}
\newblock \bibinfo{title}{{On the measurement of a weak classical force coupled
  to a quantum-mechanical oscillator. I. Issues of principle}}.
\newblock \emph{\bibinfo{journal}{Reviews of Modern Physics}}
  \textbf{\bibinfo{volume}{52}}, \bibinfo{pages}{341--392}
  (\bibinfo{year}{1980}).
\newblock \urlprefix\url{https://doi.org/10.1103/revmodphys.52.341}.

\bibitem{ZanardiRasetti1997}
\bibinfo{author}{Zanardi, P.} \& \bibinfo{author}{Rasetti, M.}
\newblock \bibinfo{title}{{Noiseless Quantum Codes}}.
\newblock \emph{\bibinfo{journal}{Physical Review Letters}}
  \textbf{\bibinfo{volume}{79}}, \bibinfo{pages}{3306--3309}
  (\bibinfo{year}{1997}).
\newblock \urlprefix\url{https://doi.org/10.1103/physrevlett.79.3306}.

\bibitem{Lidar1998}
\bibinfo{author}{Lidar, D.~A.}, \bibinfo{author}{Chuang, I.~L.} \&
  \bibinfo{author}{Whaley, K.~B.}
\newblock \bibinfo{title}{{Decoherence-Free Subspaces for Quantum
  Computation}}.
\newblock \emph{\bibinfo{journal}{Physical Review Letters}}
  \textbf{\bibinfo{volume}{81}}, \bibinfo{pages}{2594--2597}
  (\bibinfo{year}{1998}).
\newblock \urlprefix\url{https://doi.org/10.1103/physrevlett.81.2594}.

\bibitem{TicozziViola2008}
\bibinfo{author}{Ticozzi, F.} \& \bibinfo{author}{Viola, L.}
\newblock \bibinfo{title}{{Quantum Markovian Subsystems: Invariance,
  Attractivity, and Control}}.
\newblock \emph{\bibinfo{journal}{IEEE Transactions on Automatic Control}}
  \textbf{\bibinfo{volume}{53}}, \bibinfo{pages}{2048--2063}
  (\bibinfo{year}{2008}).
\newblock \urlprefix\url{https://doi.org/10.1109/tac.2008.929399}.

\bibitem{AlbertJiang2014}
\bibinfo{author}{Albert, V.~V.} \& \bibinfo{author}{Jiang, L.}
\newblock \bibinfo{title}{{Symmetries and conserved quantities in Lindblad
  master equations}}.
\newblock \emph{\bibinfo{journal}{Physical Review A}}
  \textbf{\bibinfo{volume}{89}}, \bibinfo{pages}{022118}
  (\bibinfo{year}{2014}).
\newblock \urlprefix\url{https://doi.org/10.1103/physreva.89.022118}.

\bibitem{Kraus2008}
\bibinfo{author}{Kraus, B.} \emph{et~al.}
\newblock \bibinfo{title}{{Preparation of entangled states by quantum Markov
  processes}}.
\newblock \emph{\bibinfo{journal}{Physical Review A}}
  \textbf{\bibinfo{volume}{78}}, \bibinfo{pages}{042307}
  (\bibinfo{year}{2008}).
\newblock \urlprefix\url{https://doi.org/10.1103/physreva.78.042307}.

\bibitem{Diehl2008}
\bibinfo{author}{Diehl, S.} \emph{et~al.}
\newblock \bibinfo{title}{{Quantum states and phases in driven open quantum
  systems with cold atoms}}.
\newblock \emph{\bibinfo{journal}{Nature Physics}}
  \textbf{\bibinfo{volume}{4}}, \bibinfo{pages}{878--883}
  (\bibinfo{year}{2008}).
\newblock \urlprefix\url{https://doi.org/10.1038/nphys1073}.

\bibitem{Verstraete2009}
\bibinfo{author}{Verstraete, F.}, \bibinfo{author}{Wolf, M.~M.} \&
  \bibinfo{author}{Ignacio~Cirac, J.}
\newblock \bibinfo{title}{{Quantum computation and quantum-state engineering
  driven by dissipation}}.
\newblock \emph{\bibinfo{journal}{Nature Physics}}
  \textbf{\bibinfo{volume}{5}}, \bibinfo{pages}{633--636}
  (\bibinfo{year}{2009}).
\newblock \urlprefix\url{https://doi.org/10.1038/nphys1342}.

\bibitem{TicozziViola2012}
\bibinfo{author}{Ticozzi, F.} \& \bibinfo{author}{Viola, L.}
\newblock \bibinfo{title}{{Stabilizing entangled states with quasi-local
  quantum dynamical semigroups}}.
\newblock \emph{\bibinfo{journal}{Philosophical Transactions of the Royal
  Society A: Mathematical, Physical and Engineering Sciences}}
  \textbf{\bibinfo{volume}{370}}, \bibinfo{pages}{5259--5269}
  (\bibinfo{year}{2012}).
\newblock \urlprefix\url{https://doi.org/10.1098/rsta.2011.0485}.

\bibitem{Brun2002}
\bibinfo{author}{Brun, T.~A.}
\newblock \bibinfo{title}{{A simple model of quantum trajectories}}.
\newblock \emph{\bibinfo{journal}{American Journal of Physics}}
  \textbf{\bibinfo{volume}{70}}, \bibinfo{pages}{719--737}
  (\bibinfo{year}{2002}).
\newblock \urlprefix\url{https://doi.org/10.1119/1.1475328}.

\bibitem{JacobsSteck2006}
\bibinfo{author}{Jacobs, K.} \& \bibinfo{author}{Steck, D.~A.}
\newblock \bibinfo{title}{{A straightforward introduction to continuous quantum
  measurement}}.
\newblock \emph{\bibinfo{journal}{Contemporary Physics}}
  \textbf{\bibinfo{volume}{47}}, \bibinfo{pages}{279--303}
  (\bibinfo{year}{2006}).
\newblock \urlprefix\url{https://doi.org/10.1080/00107510601101934}.

\bibitem{MisraSudarshan1977}
\bibinfo{author}{Misra, B.} \& \bibinfo{author}{Sudarshan, E. C.~G.}
\newblock \bibinfo{title}{{The Zeno's paradox in quantum theory}}.
\newblock \emph{\bibinfo{journal}{Journal of Mathematical Physics}}
  \textbf{\bibinfo{volume}{18}}, \bibinfo{pages}{756--763}
  (\bibinfo{year}{1977}).
\newblock \urlprefix\url{https://doi.org/10.1063/1.523304}.

\bibitem{FacchiPascazio2002}
\bibinfo{author}{Facchi, P.} \& \bibinfo{author}{Pascazio, S.}
\newblock \bibinfo{title}{{Quantum Zeno Subspaces}}.
\newblock \emph{\bibinfo{journal}{Physical Review Letters}}
  \textbf{\bibinfo{volume}{89}}, \bibinfo{pages}{080401}
  (\bibinfo{year}{2002}).
\newblock \urlprefix\url{https://doi.org/10.1103/physrevlett.89.080401}.

\bibitem{FacchiPascazio2008}
\bibinfo{author}{Facchi, P.} \& \bibinfo{author}{Pascazio, S.}
\newblock \bibinfo{title}{{Quantum Zeno dynamics: mathematical and physical
  aspects}}.
\newblock \emph{\bibinfo{journal}{Journal of Physics A: Mathematical and
  Theoretical}} \textbf{\bibinfo{volume}{41}}, \bibinfo{pages}{493001}
  (\bibinfo{year}{2008}).
\newblock \urlprefix\url{https://doi.org/10.1088/1751-8113/41/49/493001}.

\bibitem{Pallister2018}
\bibinfo{author}{Pallister, S.}, \bibinfo{author}{Linden, N.} \&
  \bibinfo{author}{Montanaro, A.}
\newblock \bibinfo{title}{{Optimal Verification of Entangled States with Local
  Measurements}}.
\newblock \emph{\bibinfo{journal}{Physical Review Letters}}
  \textbf{\bibinfo{volume}{120}}, \bibinfo{pages}{170502}
  (\bibinfo{year}{2018}).
\newblock \urlprefix\url{https://doi.org/10.1103/physrevlett.120.170502}.

\bibitem{ZhuHayashi2019Adversarial}
\bibinfo{author}{Zhu, H.} \& \bibinfo{author}{Hayashi, M.}
\newblock \bibinfo{title}{{Efficient Verification of Pure Quantum States in the
  Adversarial Scenario}}.
\newblock \emph{\bibinfo{journal}{Physical Review Letters}}
  \textbf{\bibinfo{volume}{123}}, \bibinfo{pages}{260504}
  (\bibinfo{year}{2019}).
\newblock \urlprefix\url{https://doi.org/10.1103/physrevlett.123.260504}.

\bibitem{Yu2022}
\bibinfo{author}{Yu, X.-D.}, \bibinfo{author}{Shang, J.} \&
  \bibinfo{author}{G{\"{u}}hne, O.}
\newblock \bibinfo{title}{{Statistical Methods for Quantum State Verification
  and Fidelity Estimation}}.
\newblock \emph{\bibinfo{journal}{Advanced Quantum Technologies}}
  \textbf{\bibinfo{volume}{5}}, \bibinfo{pages}{2100126}
  (\bibinfo{year}{2022}).
\newblock \urlprefix\url{https://doi.org/10.1002/qute.202100126}.

\bibitem{Page1993}
\bibinfo{author}{Page, D.~N.}
\newblock \bibinfo{title}{{Average entropy of a subsystem}}.
\newblock \emph{\bibinfo{journal}{Physical Review Letters}}
  \textbf{\bibinfo{volume}{71}}, \bibinfo{pages}{1291--1294}
  (\bibinfo{year}{1993}).
\newblock \urlprefix\url{https://doi.org/10.1103/physrevlett.71.1291}.

\bibitem{AlMohyHigham2011}
\bibinfo{author}{Al-Mohy, A.~H.} \& \bibinfo{author}{Higham, N.~J.}
\newblock \bibinfo{title}{{Computing the Action of the Matrix Exponential, with
  an Application to Exponential Integrators}}.
\newblock \emph{\bibinfo{journal}{SIAM Journal on Scientific Computing}}
  \textbf{\bibinfo{volume}{33}}, \bibinfo{pages}{488--511}
  (\bibinfo{year}{2011}).
\newblock \urlprefix\url{https://doi.org/10.1137/100788860}.

\bibitem{Harris2020}
\bibinfo{author}{Harris, C.~R.} \emph{et~al.}
\newblock \bibinfo{title}{{Array programming with NumPy}}.
\newblock \emph{\bibinfo{journal}{Nature}} \textbf{\bibinfo{volume}{585}},
  \bibinfo{pages}{357--362} (\bibinfo{year}{2020}).
\newblock \urlprefix\url{https://doi.org/10.1038/s41586-020-2649-2}.

\bibitem{Virtanen2020}
\bibinfo{author}{Virtanen, P.} \emph{et~al.}
\newblock \bibinfo{title}{{SciPy 1.0: fundamental algorithms for scientific
  computing in Python}}.
\newblock \emph{\bibinfo{journal}{Nature Methods}}
  \textbf{\bibinfo{volume}{17}}, \bibinfo{pages}{261--272}
  (\bibinfo{year}{2020}).
\newblock \urlprefix\url{https://doi.org/10.1038/s41592-019-0686-2}.

\bibitem{Hunter2007}
\bibinfo{author}{Hunter, J.~D.}
\newblock \bibinfo{title}{{Matplotlib: A 2D Graphics Environment}}.
\newblock \emph{\bibinfo{journal}{Computing in Science \& Engineering}}
  \textbf{\bibinfo{volume}{9}}, \bibinfo{pages}{90--95} (\bibinfo{year}{2007}).
\newblock \urlprefix\url{https://doi.org/10.1109/mcse.2007.55}.

\bibitem{Lean2021}
\bibinfo{author}{de~Moura, L.} \& \bibinfo{author}{Ullrich, S.}
\newblock \bibinfo{title}{The {Lean 4} theorem prover and programming
  language}.
\newblock In \emph{\bibinfo{booktitle}{Automated Deduction -- CADE 28}},
  \bibinfo{pages}{625--635} (\bibinfo{publisher}{Springer},
  \bibinfo{year}{2021}).
\newblock \urlprefix\url{https://doi.org/10.1007/978-3-030-79876-5_37}.

\bibitem{Mathlib2020}
\bibinfo{author}{{The mathlib Community}}.
\newblock \bibinfo{title}{The {Lean} mathematical library}.
\newblock In \emph{\bibinfo{booktitle}{Proceedings of the 9th ACM SIGPLAN
  International Conference on Certified Programs and Proofs}},
  \bibinfo{pages}{367--381} (\bibinfo{publisher}{ACM}, \bibinfo{year}{2020}).
\newblock \urlprefix\url{https://doi.org/10.1145/3372885.3373824}.

\end{thebibliography}
\end{document}

\maketitle\vspace{-2em}
This supplement supplies finite-dimensional proofs, an explicit measurement instrument, numerical conventions and the boundaries of the scar interpretation. Unless stated otherwise, $\hbar=1$, $N\geq5$, all Hilbert spaces are finite dimensional, and all Hamiltonians denoted $H$ are Hermitian. The symbol $P$ denotes the rank-one cluster projector, $R=\id-P$ its complement, and $Q_i$ a local syndrome-defect projector. These are distinct objects.

\section{What local measurement data can establish}
\subsection{A fixed product-basis obstruction}
Let $\mathcal B=\{\ket{b}=\bigotimes_i\ket{b_i}\}$ be a fixed orthonormal product basis. Its complete projective measurement has effects $\Pi_b=\ket{b}\bra b$. For an arbitrary density matrix $\rho$, define
\begin{equation}
\Delta_{\mathcal B}(\rho)=\sum_b\Pi_b\rho\Pi_b
=\sum_b\tr(\Pi_b\rho)\bigotimes_i\ket{b_i}\bra{b_i}.
\label{eq:dephase}
\end{equation}
The second expression is a convex mixture of product states, hence fully separable. Yet
\begin{equation}
\tr[\Pi_b\Delta_{\mathcal B}(\rho)]=\tr(\Pi_b\rho)
\end{equation}
for every $b$. All marginals, correlators diagonal in $\mathcal B$, and deterministic or randomized classical processing of those outcomes therefore have the same probability distribution for $\rho$ and this separable explanation. This is an information-theoretic obstruction for unrestricted entanglement certification. It does not preclude supervised recognition under an ensemble promise, such as the local-fingerprint task in Ref.~\cite{Li2026Fingerprint}.

The same formula gives the disturbance of actual factor-resolved readout. Conditioned on a complete outcome, the postmeasurement state is the product state $\ket b$; if outcomes are discarded, it is Eq.~\eqref{eq:dephase}. Post-processing outcomes into a parity cannot undo the extra quantum dephasing already produced. For a ring cluster measured in the all-$Z$ basis, all $2^N$ probabilities equal $2^{-N}$, so $\Delta_Z(P)=\id/2^N$ and its cluster fidelity is $2^{-N}$. This is directly distinguishable from a parity-only instrument.

Let $K_i=Z_{i-1}X_iZ_{i+1}$ and $\Pi_{i,\pm}=(\id\pm K_i)/2$. A coarse two-outcome measurement produces $\Pi_{i,\pm}\rho\Pi_{i,\pm}$ and satisfies $\Pi_{i,+}P\Pi_{i,+}=P$, $\Pi_{i,-}P\Pi_{i,-}=0$. The cluster state is exactly preserved. Stabilizer witnesses and verification protocols use related algebra,\cite{TothGuhne2005,GuhneToth2009,Dangniam2020} but the measurement instrument determines whether the measured copy remains useful. Nondemolition implementations are already established.\cite{Liu2021QND,Riera2023}

\subsection{A simple fidelity and entanglement certificate}
The defect spectrum below implies $R\leq D\leq NR$. Taking expectation values gives
\begin{equation}
\max\{0,1-\tr(D\rho)\}\leq\tr(P\rho)
\leq 1-\frac1N\tr(D\rho).
\label{eq:witness}
\end{equation}
Estimating local defects on independently prepared copies can thus provide an input fidelity lower bound. This procedure is not itself a claim that an arbitrary subsequently tested copy has the same state; source assumptions must be stated.

For a connected graph, across every nontrivial bipartition the graph state has at least two equal nonzero Schmidt coefficients. Its largest squared Schmidt coefficient is at most $1/2$. A pure state product across that partition has squared overlap at most $1/2$, by the variational characterization of the largest Schmidt coefficient. Convexity extends the bound to mixtures of states separable across potentially different cuts. Consequently $\tr(P\rho)>1/2$ certifies genuine multipartite entanglement. The ring graph has this property. Graph-state entanglement and Clifford transformations are discussed in Refs.~\cite{Hein2004,VanDenNest2004}; here the argument only requires the Schmidt bound.

\section{Exact embedding, locality and algebraic sectors}
\subsection{Graph basis and defect spectrum}
For a ring let $U=\prod_i\mathrm{CZ}_{i,i+1}$ and $W=U\mathsf h^{\otimes N}$. Since $\mathsf h Z\mathsf h=X$ and $UX_iU^\dagger=K_i$, one has $WZ_iW^\dagger=K_i$. Write $\ket{C_s}=W\ket s$, $s\in\{0,1\}^N$. Then
\begin{equation}
K_i\ket{C_s}=(-1)^{s_i}\ket{C_s},\qquad Q_i\ket{C_s}=s_i\ket{C_s},\qquad
D\ket{C_s}=|s|\ket{C_s}.
\end{equation}

The transformation is unitary, so this is a complete orthonormal eigenbasis. There is one zero-defect vector, $\ket{C_0}=\Cstate$. Therefore
\begin{equation}
P=\prod_i(\id-Q_i),\qquad R\leq D\leq NR.
\label{eq:Dspectrum}
\end{equation}
This proof establishes a size-independent lower eigenvalue at fixed per-generator weight, not a size-independent hardware cost.

\subsection{Hamiltonian and support}
In the syndrome frame consider
\begin{equation}
\widetilde H=J\sum_{i,\delta=\pm1}n_i(a_{i\delta}X_j+b_{i\delta}Y_j+c_{i\delta}Z_j),\qquad j=i+\delta,
\quad n_i=(\id-Z_i)/2.
\end{equation}
The factors act on distinct sites, so each term is Hermitian and annihilates the vacuum. Consequently $HP=PH=0$ for $H=W\widetilde HW^\dagger$. Its physical form is
\begin{equation}
H=J\sum_{i,\delta=\pm1}Q_i\left[a_{i\delta}Z_j-b_{i\delta}Z_{j-1}Y_jZ_{j+1}+c_{i\delta}K_j\right].
\label{eq:physicalH}
\end{equation}
$Q_i$ commutes with the entire bracket because its preimage $n_i$ commutes with all operators on $j$. The union of the supports is contained in four consecutive sites. Apparent Pauli anticommutations on individual overlapping sites cancel in the complete product. The model is local even though $W$ is a many-qubit unitary.

Each syndrome-frame term has zero trace, because $\tr(n_i\sigma_j)=\tr(n_i)\tr(\sigma_j)$ times the dimension of spectator sites and $\tr\sigma_j=0$. Thus $\tr H=0$. The canonical infinite-temperature energy is exactly the target energy. This observation fixes a thermal reference but does not replace a microcanonical finite-size analysis or prove ensemble equivalence.

\subsection{Connectivity outside the target}
For generic coefficients, a bit can flip if either neighbour is occupied. Given any nonempty configuration, select an occupied site and successively fill neighbouring zeros around the ring. Every step is allowed, reaching the all-one configuration. Since $H$ is Hermitian, the transition edges are undirected, so any two nonempty configurations are connected through the all-one configuration. No move connects the vacuum to this component.

In a particular sample, contributions from two occupied neighbours can in principle cancel. With independently sampled continuous coefficients, exact cancellation on any prescribed edge is a measure-zero event. A fixed finite ring has finitely many edges, so the probability of any such cancellation remains zero. The separately supplied structural check constructs the graph for all 51 realized Hamiltonians and finds component sizes $1$ and $2^N-1$ for each. Graph connectivity rules out computational-basis fragmentation beyond the isolated vacuum; it does not alone prove quantum chaos.

\subsection{Commutant of the independent term family}
Suppose an operator $T$ commutes with every independent term $n_iX_j,n_iY_j,n_iZ_j$ on every directed nearest-neighbour bond. Since $(n_iX_j)^2=n_i$, it also commutes with every $n_i$. The joint eigenspaces of all $n_i$ are one-dimensional computational basis states, so $T$ is diagonal in that basis. Commutation with a nonzero flip matrix element enforces equal diagonal values at the two ends of the edge. Connectivity therefore enforces one common value on all nonempty bitstrings, and permits a separate value on the vacuum. Conversely both vacuum and complementary projectors commute with the family. The common commutant is exactly their linear span; graph dressing yields $\operatorname{span}\{P,R\}$.

This is a statement about the entire independently parametrized term family. A fixed Hamiltonian always commutes with its spectral projectors, and could have additional coefficient-dependent structure. The algebraic result must not be promoted to a universal ETH theorem.

\subsection{Antiunitary symmetry class}
On a fixed directed bond define $A=n_iX_j$, $B=n_iY_j$ and $C=n_iZ_j$. They obey $[A,B]=2iC$. If one coefficient-independent antiunitary operator fixed all members of the family, it would fix $A,B,C$. Acting on the commutator would instead give $[A,B]=-2iC$, a contradiction. Thus there is no common antiunitary fixing all independently variable terms. This supports the GUE reference for the random ensemble. It does not prove random-matrix universality.

For a more limited numerical test, the phase of a product of hopping amplitudes around a closed configuration-space loop is invariant under diagonal basis rephasing. A loop product with nonzero imaginary part excludes making that sample real by diagonal rephasing. The structural audit finds such loops for every sampled Hamiltonian. This check does not exclude every conceivable nonlocal, coefficient-dependent antiunitary transformation, and no such exclusion is needed for the filtering proof.

\subsection{Area-law entropy of the target}
For a graph state and a bipartition $A|B$, the Schmidt rank is $2^{\operatorname{rank}_{\mathbb F_2}\Gamma_{AB}}$, with equal Schmidt coefficients. For the half-ring cut at $N\geq5$, two distinct edges cross the cut with independent endpoint incidence, so the binary adjacency rank is two. Hence
\begin{equation}
S_A(P)=2\ln2.
\end{equation}
One can alternatively obtain this by reducing the two crossing controlled-phase gates to two independent Bell-pair Schmidt factors using local unitaries within $A$ and $B$. Internal graph edges do not change the bipartite entropy. The $N=6$ complementary mean entropy in the reported sample is lower than $2\ln2$; thus we do not claim a low-entropy outlier at every small size. The separation becomes evident at larger sizes in the computed sequence.

\section{Perfect-detection filtering for arbitrary mixed states}
\begin{theorem}[Finite-time cluster filtering]
Assume $H=H^\dagger$, $HP=PH=0$, $Q_iP=PQ_i=0$, and Eq.~\eqref{eq:Dspectrum}. Let $\gamma>0$, $A(t)=e^{(-iH-\gamma D/2)t}$ and $\rho_0$ be any density matrix. Put $p=\tr(P\rho_0)$. Then $\tr(PA\rho_0A^\dagger)=p$ and
\begin{align}
S(t)&=\tr(A\rho_0A^\dagger)=p+q(t),\\
(1-p)e^{-N\gamma t}&\leq q(t)\leq(1-p)e^{-\gamma t},\\
\Fid(t)&=p/S(t).
\label{eq:mixedtheorem}
\end{align}
At any finite time $A(t)$ is invertible, so $S(t)>0$. If $p>0$, $\Fid(t)\to1$ and $S(t)\to p$. If $p=0$, $\Fid(t)=0$ at every finite time.
\end{theorem}
\begin{proof}
The reducing projector identities imply $A=P+B$ with $B=RAR$. For $x\in R$, let $y(t)=B(t)x$. Hermiticity cancels the Hamiltonian from the norm derivative:
\begin{equation}
\frac{\mathrm d}{\mathrm dt}\|y\|^2
=\bra y(iH-\gamma D/2-iH-\gamma D/2)\ket y
=-\gamma\bra yD\ket y.
\end{equation}
Since $y\in R$, Eq.~\eqref{eq:Dspectrum} yields
\begin{equation}
-N\gamma\|y\|^2\leq\frac{\mathrm d}{\mathrm dt}\|y\|^2\leq-\gamma\|y\|^2.
\end{equation}
Integrating gives the two exponential inequalities and, equivalently,
\begin{equation}
e^{-N\gamma t}R\leq B^\dagger B\leq e^{-\gamma t}R.
\end{equation}
The positive operator $\rho_R=R\rho_0R$ has trace $1-p$, so multiplication by $\rho_R$ and taking the trace gives the stated bounds on $q=\tr(B\rho_RB^\dagger)$. Expanding $(P+B)\rho_0(P+B)^\dagger$ shows that the target block is $pP$. Target--bright coherence blocks are allowed; their traces vanish. This proves $S=p+q$ and $\Fid=p/S$. The limits and zero-overlap case follow immediately.
\end{proof}

The proof uses the dissipative part as a positive quadratic form. It neither diagonalizes $\Heff$ nor assumes orthogonal right eigenvectors, a complete eigenbasis or the absence of exceptional points. The operator-norm bound $\|B(t)\|\leq e^{-\gamma t/2}$ is stronger than an imaginary-part spectral bound for controlling transients.

For a right eigenvector $v\in R$ with $\Heff v=zv$, multiplication by $v^\dagger$ gives
\begin{equation}
\operatorname{Im}z=-\frac\gamma2\frac{\bra vD\ket v}{\|v\|^2}\leq-\gamma/2.
\end{equation}
The exact target eigenvalue is zero. No exceptional-point or non-Hermitian topological claim is inferred from the plot. General non-Hermitian dynamics and heralding have broader settings,\cite{Ashida2020,LeeChan2014} but their extra structures are not required here.

\section{Inefficient detection, nonuniform rates and dark counts}
Let $\gamma_i>0$ and $0\leq\eta_i\leq1$. Split each physical jump into an observed channel $\sqrt{\eta_i\gamma_i}Q_i$ and an unobserved channel $\sqrt{(1-\eta_i)\gamma_i}Q_i$. Conditional on no observed events, the unnormalized state follows
\begin{equation}
\dot{\widetilde\rho}=-i[H,\widetilde\rho]
+\sum_i(1-\eta_i)\gamma_iQ_i\widetilde\rho Q_i
-\frac12\{G,\widetilde\rho\},\qquad G=\sum_i\gamma_iQ_i.
\label{eq:inefficient}
\end{equation}
This equation is a completely positive trace-decreasing evolution. Its target block remains $pP$. The complementary block remains positive, while
\begin{equation}
\dot q=-\tr(G_{\rm obs}\widetilde\rho_R),\qquad
G_{\rm obs}=\sum_i\eta_i\gamma_iQ_i.
\end{equation}
Define $g_{\rm obs}=\min_i\eta_i\gamma_i$ and $G_{\rm obs,tot}=\sum_i\eta_i\gamma_i$. In the syndrome basis the smallest bright eigenvalue of $G_{\rm obs}$ is exactly $g_{\rm obs}$, so
\begin{equation}
-G_{\rm obs,tot}q\leq\dot q\leq-g_{\rm obs}q.
\end{equation}
Integration proves
\begin{equation}
(1-p)e^{-G_{\rm obs,tot}t}\leq q(t)\leq(1-p)e^{-g_{\rm obs}t},\qquad
S=p+q,\quad \Fid=p/S.
\label{eq:effbound}
\end{equation}
Uniform rates and efficiencies recover the replacements $\gamma\mapsto\eta\gamma$ and $N\gamma\mapsto N\eta\gamma$. If some $\eta_i=0$, this simple lower-gap guarantee becomes zero. A specific Hamiltonian might still move defects into detectable configurations, but that requires a separate observability bound. No positive universal rate is asserted in that case.

For uniform efficiency and fixed $\Gamma=\sum_i\gamma_i$, one has $g=\min_i\gamma_i\leq\Gamma/N$, with equality precisely for uniform allocation. This maximizes the guaranteed exponent among the fixed generator checks. Tightness follows from $H=0$ and a single defect on a minimum-rate site, which gives $q(t)=q(0)e^{-\eta\gamma_{\min}t}$. Thus the rate optimization is a minimax guarantee over the allowed Hamiltonian class, not an optimization theorem for the actual noncommuting benchmark Hamiltonian.

State-independent detector dark counts at total Poisson rate $\nu$ multiply every no-observed-click map by $e^{-\nu t}$. They decrease success to $e^{-\nu t}S(t)$ without changing the normalized retained state, provided they do not act on the system and their rate is independent of its state. State-dependent false counts or misidentified outcomes require different jump operators and are not covered by that simplification.

\section{An explicit target-preserving measurement instrument}
A discrete weak measurement of one defect projector can be defined exactly by
\begin{equation}
M_{i,0}=\id-Q_i+\sqrt{1-\kappa_i}\,Q_i,\qquad
M_{i,1}=\sqrt{\kappa_i}\,Q_i,\qquad 0\leq\kappa_i\leq1.
\end{equation}
Orthogonality of $Q_i$ and $\id-Q_i$ implies $M_{i,0}^\dagger M_{i,0}+M_{i,1}^\dagger M_{i,1}=\id$. An ancilla unitary realizing this measurement is
\begin{equation}
U_i=(\id-Q_i)\otimes\id_a+Q_i\otimes e^{-i\theta_iY_a},
\qquad \theta_i=\arcsin\sqrt{\kappa_i}\in[0,\pi/2],
\end{equation}
with the ancilla initialized in $\ket0_a$ and measured in its computational basis. Since $e^{-i\theta Y}\ket0=\cos\theta\ket0+\sin\theta\ket1$, the two Kraus operators above follow. The coupling is to the stabilizer parity as a whole, not to three independently recorded local Pauli outcomes. The construction is an ideal instrument; a compiled hardware circuit must be separately calibrated.

Apply $e^{-iH\Delta t}$ and the weak checks with $\kappa_i=\gamma_i\Delta t$ in each sufficiently small step. Then
\begin{equation}
M_{i,0}=\id-\tfrac12\gamma_i\Delta t\,Q_i+O(\Delta t^2),
\end{equation}
and the all-no-click product has generator $-iH-\frac12\sum_i\gamma_iQ_i$. Keeping the click branches produces the Lindblad equation. This limit is controlled for every fixed finite $N$ as $\Delta t\to0$; at finite step size commutator and multiple-event errors must be assessed. The reported simulations integrate the limiting equations directly and do not assume a finite-step circuit is exact. Quantum-trajectory derivations are standard.\cite{Dum1992,Molmer1993,Dalibard1992,PlenioKnight1998}

The projector $P$ commutes with both $H$ and the ancilla coupling, so its binary value is a conserved nondemolition observable. Each $K_i$ commutes with the parity measurements but not, generally, with $H$. If one imposed $[H,K_i]=0$ for all independent $K_i$, $H$ would be diagonal in their complete graph basis. That is a different, commuting-stabilizer setting and does not furnish the chaotic bright dynamics tested here.

For the unconditional generator,
\begin{equation}
\mathcal L(\rho)=-i[H,\rho]+\gamma\sum_i\left(Q_i\rho Q_i-\tfrac12\{Q_i,\rho\}\right),
\end{equation}
$\mathcal L^\dagger(P)=0$ and $\mathcal L(\id)=0$. Because $H$ reduces $R$ and $Q_iRQ_i=Q_i$, one also has $\mathcal L(R)=0$. Hence $P$, $R/(2^N-1)$, and every convex combination are stationary. In particular, the target is not the unique attractive state. More general dissipative state-engineering constructions can have different attractivity properties.\cite{Kraus2008,TicozziViola2012,Wang2024DFS}

\section{A finite-time bound for coherent Hamiltonian errors}
Let $H'=H+V$, with $V=V^\dagger$, and assume ideal detection with uniform $\gamma>0$. Start from $\Cstate$. Decompose the unnormalized no-click vector as
\begin{equation}
\ket{\widetilde\psi(t)}=a(t)\Cstate+\ket{b(t)},\qquad b(t)\in R,\quad a(0)=1,\quad b(0)=0.
\end{equation}
Define $v_0=\bra{C_N}V\ket{C_N}$, $w=RV\Cstate$, $v=\|w\|=\|RVP\|$ and $\alpha=\gamma/2$. The equations are
\begin{align}
\dot a&=-iv_0a-i\bra w b\rangle,\\
\dot b&=\left[-iR(H+V)R-\gamma D_R/2\right]b-iwa.
\end{align}
The phase $v_0$ can be removed. The block $RVR$ only changes the Hermitian bright Hamiltonian, so its semigroup retains norm bound $e^{-\alpha t}$. The whole no-click evolution is contractive, giving $|a(t)|\leq1$. Duhamel's formula yields
\begin{equation}
\|b(t)\|\leq v\int_0^t e^{-\alpha(t-s)}|a(s)|\,\mathrm ds
\leq\frac v\alpha(1-e^{-\alpha t})=:b_*(t).
\end{equation}
For $\bar a=e^{iv_0t}a$, $|\dot{\bar a}|\leq v\|b\|$, and thus
\begin{equation}
|a(t)|\geq1-v\int_0^t b_*(s)\,\mathrm ds
=1-\frac{v^2}{\alpha}\left[t-\frac{1-e^{-\alpha t}}\alpha\right].
\end{equation}
Let $a_*$ be the positive part of the right side. If $a_*>0$, monotonicity in the nonnegative numerator and denominator components gives
\begin{equation}
\frac{|a|^2}{|a|^2+\|b\|^2}\geq\frac{a_*^2}{a_*^2+b_*^2},\qquad
S=|a|^2+\|b\|^2\geq a_*^2.
\end{equation}
A negative amplitude lower bound must never be squared to create a spurious fidelity guarantee. Once the un-clipped amplitude lower bound is nonpositive, the nontrivial guarantee has expired. The numerical code implements this clipping explicitly.

For the physical perturbation $V=vZ_0$, $W^\dagger Z_0W=X_0$ and $\|RV P\|=|v|$, so the input parameter of the bound equals the simulated error amplitude. The curves use $v/J=0.2$. They do not cover incoherent jumps, finite efficiency, a distribution of calibration errors, or an arbitrary imperfect initial state. No long-time perfect-protection claim follows from this finite-time estimate.

\section{Soundness, sampling and preparation cost}
\subsection{An orthogonal input and a promised input}
For $p=0$, Eq.~\eqref{eq:effbound} gives $S(t)\leq e^{-g_{\rm obs}t}$. This is a false-acceptance probability bound for an orthogonal state under the stated detector model. The normalized state in such a rare accepted branch still has zero cluster fidelity. Acceptance is therefore not a logical proof of the target state in a single run.

If an independent promise gives $p\geq p_{\min}>0$, then
\begin{equation}
\Fid(t)\geq\frac{p_{\min}}{p_{\min}+(1-p_{\min})e^{-g_{\rm obs}t}}.
\end{equation}
The distinction between the two guarantees is essential: one bounds type-I acceptance for an alternative input, while the other bounds conditional fidelity under a preparation promise.

\subsection{Independent-copy verification}
For an identically prepared independent source whose fidelity is at most $1-\epsilon$, the single-test acceptance probability is at most
\begin{equation}
1-\epsilon(1-e^{-g_{\rm obs}t}).
\end{equation}
The probability of accepting all $m$ independent tests is at most the $m$th power. A significance bound $\delta$ is therefore attained if
\begin{equation}
m\geq\frac{\log\delta}{\log[1-\epsilon(1-e^{-g_{\rm obs}t})]}.
\end{equation}
Here $0<\epsilon,\delta<1$ and $g_{\rm obs}t>0$ are assumed; if $g_{\rm obs}t=0$, this bound gives no finite sample guarantee. Round the expression up to an integer. Independence, stationarity of the source and trusted detector behaviour are assumptions, not outcomes of the bound. An adversarial or correlated source needs an appropriate sampling protocol.\cite{ZhuHayashi2019Adversarial,Yu2022} Direct fidelity estimation and local verification provide established alternatives.\cite{FlammiaLiu2011,Pallister2018,HayashiMorimae2015,TakeuchiMorimae2018,ZhuHayashi2019Hypergraph,Li2023}

The finite-trial plot uses only the tight $H=0$, one-defect case, for which the first-click waiting time is exponential with rate $\gamma$. At each of 33 times there are 20,000 independent Bernoulli draws with probability $e^{-\gamma t}$ and fixed seed 271828. Wilson intervals are computed with $z=1.959963984540054$. They are pointwise nominal 95\% intervals, not exact finite-sample coverage guarantees or a simultaneous band; the points are synthetic Monte Carlo data, not quantum hardware observations.

\subsection{Resource lower bounds}
Assume no Hamiltonian errors or dark counts, $p>0$, and $g_{\rm obs}>0$. Then $\Fid(t)S(t)=p$. If $\Fid(t)\geq1-\epsilon_f$, then
\begin{equation}
S(t)\leq\frac{p}{1-\epsilon_f},\qquad
\mathbb E[\text{independent attempts}]\geq\frac{1-\epsilon_f}{p}.
\end{equation}
As $t\to\infty$, the mean attempt count tends to $1/p$. This assumes each failed attempt is replaced by the same freshly prepared input; it excludes adaptive recycling or feedback.

For independent physical $Z$ errors with probability $r$, the input is diagonal in the graph basis and the exact target population is $(1-r)^N$. For a maximally mixed input it is $2^{-N}$. A Haar-random pure state has this latter overlap only on average, not deterministically. The plotted curves use the two explicit mixed-state families, avoiding a hidden typicality assumption.

With $H=0$, independent phase-error inputs give the additional closed form
\begin{equation}
S(t)=\left(1-r+r e^{-\eta\gamma t}\right)^N,\qquad
\Fid(t)=\left(\frac{1-r}{1-r+r e^{-\eta\gamma t}}\right)^N.
\end{equation}
The universal bound is generally more conservative because it allows arbitrary bright inputs and Hamiltonians. No strict advantage over this commuting baseline is inferred from a common bound.

Fixed per-site rate and fixed total rate are different limits. If $p$ decreases exponentially with $N$, the sufficient time contains a term proportional to $N$ even at fixed $\gamma$. At fixed total $\Gamma$, the additional $N/\Gamma$ factor can make the sufficient-time estimate quadratic. These are upper bounds on a sufficient duration for the specified promise, not universal lower bounds on the actual mixing time of the chosen interacting model. The preparation-attempt obstruction, by contrast, follows exactly from $\Fid S=p$.

\section{Subspaces, coherence and the limits of generalization}
For a subspace projector $P_{\mathcal S}$ and $R_{\mathcal S}=\id-P_{\mathcal S}$, no-click invariance is equivalent to
\begin{equation}
R_{\mathcal S}\Heff P_{\mathcal S}=0.
\end{equation}
Necessity follows by differentiating the propagator at zero; sufficiency follows by exponentiation of its block-triangular form. Invariance under the full Lindblad evolution additionally requires $R_{\mathcal S}L_iP_{\mathcal S}=0$ for each jump. To see necessity, the initial rate of population leaving a pure state in the subspace is $\sum_i\|R_{\mathcal S}L_i\psi\|^2$. Once these terms vanish, the off-diagonal generator condition reduces to the no-click invariance condition. These are standard Markovian-subspace criteria.\cite{TicozziViola2008}

Under invariance, unitary dynamics for every encoded state requires each restricted jump to be a scalar: $L_iP_{\mathcal S}=\lambda_iP_{\mathcal S}$. The purity derivative for a pure state is
\begin{equation}
\frac{\mathrm d}{\mathrm dt}\tr\rho^2=-2\sum_i\left(\langle L_i^\dagger L_i\rangle-|\langle L_i\rangle|^2\right).
\end{equation}
Every nonnegative variance must vanish; linearity on superpositions forces the same scalar on the entire subspace. Distinct measurement eigenvalues can preserve individual basis states while destroying their coherences. Thus the rank-one result in this paper does not automatically protect a scar tower or an encoded quantum memory. Algebraic tower constructions and topological scars address different additional structures.\cite{Choi2019,Mark2020,Ok2019}

For example, if two stationary vectors have real measurement eigenvalues $\lambda_1\ne\lambda_2$ for one Hermitian jump, their off-diagonal density element decays at rate $(\lambda_1-\lambda_2)^2/2$. Both populations can remain unchanged while the coherent superposition is lost. Claiming preservation of an entire tower requires checking the common-scalar condition, not only the survival of each eigenvector.

\section{Exact numerical specification and checks}
\subsection{Data generation}
The code uses the syndrome basis for sparse construction and dynamics, and transforms eigenvectors into the physical basis before calculating entanglement. Matrix elements of a Pauli $Y$ satisfy $Y\ket0=i\ket1$, $Y\ket1=-i\ket0$; this sign convention is independently checked. All random coefficients and initial-state seeds are fixed before spectral calculations. Table~\ref{tab:sizes} specifies every Hermitian spectral realization.

\begin{table}[htbp]\centering
\caption{Hermitian spectral calculations. Seeds run from $91000+100N$ through that value plus the realization count minus one. The central half of each complementary spectrum is used for gap ratios.}
\label{tab:sizes}
\begin{tabular}{rrrrr}\toprule
$N$ & Complement dimension & Realizations & Mean $r$ & Target entropy\\\midrule
6 &63 &12 &0.542153 &$2\ln2$\\
7 &127 &12 &0.566350 &$2\ln2$\\
8 &255 &10 &0.561400 &$2\ln2$\\
9 &511 &8 &0.576604 &$2\ln2$\\
10 &1023 &6 &0.590822 &$2\ln2$\\
11 &2047 &3 &0.595782 &$2\ln2$\\\bottomrule
\end{tabular}\end{table}

For complementary eigenvalues $E_1<\cdots<E_d$, define $\delta_j=E_{j+1}-E_j$ and $r_j=\min(\delta_j,\delta_{j+1})/\max(\delta_j,\delta_{j+1})$. No unfolding is applied. The code selects eigenvalue indices from $\lfloor0.25d\rfloor$ to $\lfloor0.75d\rfloor-1$, then forms gaps and ratios inside that slice. The target is removed exactly as a one-dimensional invariant block, not by deleting the eigenvalue nearest zero from a numerical list. No accidental small bright eigenvalues are deleted. Random-matrix reference values are comparison benchmarks,\cite{Atas2013} not fitted targets.

For the entropy summary, only the first coefficient realization per size is used. Select the 64 bright eigenstates closest to zero; at $N=6$ only 63 exist. The mean and sample standard deviation summarize those eigenstates. They are neither an ensemble mean over all random Hamiltonians nor a confidence interval. The $N=11$ bootstrap interval for level statistics is based on just three realization means and must not be interpreted as a precise asymptotic estimate.

\subsection{State and density-matrix propagation}
Pure conditional dynamics use $N=10$, $p\in\{0.05,0.5\}$, $\gamma/J\in\{0.25,1,4\}$, and 241 evenly spaced times over $0\leq Jt\leq12$. The displayed figure uses $p=0.05$, while both datasets are supplied. The pure initial vector is $\sqrt p\ket0+\sqrt{1-p}\ket{v_R}$, where $v_R$ is the normalized complex Gaussian complementary vector from seed 404.

The inefficient-detection dataset uses $N=6$, $p=0.2$, state seed 405, $\gamma/J=1$, $\eta\in\{0,0.25,0.5,1\}$ and 121 times over $0\leq Jt\leq8$. With column-major vectorization,
\begin{equation}
\operatorname{vec}(AXB)=(B^T\otimes A)\operatorname{vec}(X),
\end{equation}
so the Hamiltonian superoperator is $-i(\id\otimes H-H^T\otimes\id)$. Projectors $n_i$ are real and diagonal in this frame, giving the superoperator used by the code. The $\eta=1$ density evolution is separately compared with an outer product of the no-click wavefunction; $\eta=0$ is checked for trace preservation and constant target weight.

The coherent-error dataset uses $N=8$, initial target, physical $V=0.2JZ_0$, $\gamma/J\in\{0,1,4,16\}$ and 301 times over $0\leq Jt\leq30$. The complex-spectrum calculation uses $N=8$, $\gamma/J=1$ and no error. The largest interacting Hermitian calculation is $N=11$; the $N=100$ curves are explicit scalar resource formulas only.

\subsection{Independent implementations}
The independent check builds the $N=5$ Hamiltonian by dense Kronecker products of Pauli matrices, rather than sparse bit flips. It separately constructs the Clifford and stabilizers, verifies their conjugation identities, integrates the matrix differential equation directly, and compares with the sparse superoperator. The principal measured discrepancies are given in Table~\ref{tab:checks}.

\begin{table}[htbp]\centering
\caption{Independent numerical checks. Frobenius norms are used unless otherwise specified. Values are rounded upward or to the displayed precision. These are floating-point discrepancies, not experimental uncertainties.}
\label{tab:checks}
\begin{tabular}{p{0.70\linewidth}r}\toprule
Check & Discrepancy\\\midrule
Sparse bit construction versus dense Kronecker $\widetilde H$ & $0$\\
Maximum $Wn_iW^\dagger-Q_i$ & $1.79\times10^{-15}$\\
Physical target residual $\|H\Cstate\|$ & $4.17\times10^{-16}$\\
Direct density ODE versus sparse generator & $\leq5.09\times10^{-13}$\\
Perfect-detector density versus wavefunction outer product & $8.35\times10^{-17}$\\
Product-basis probabilities before/after dephasing, 243 bases & $\leq2.36\times10^{-16}$\\\bottomrule
\end{tabular}\end{table}

The full simulation also records Hermiticity, eigensystem residuals, density-matrix positivity to numerical tolerance, the complex spectral gap and target-population conservation. The independent CSV verifier checks the identities and bounds throughout the supplied time grids. Verification code, raw numerical reports and plotting inputs are included, so each displayed point can be regenerated. NumPy, SciPy and Matplotlib provide the array, linear algebra and plotting implementations.\cite{Harris2020,Virtanen2020,Hunter2007,AlMohyHigham2011}

\subsection{Additional mixed-state and instrument validation}
All nine original numerical CSV files were regenerated with the original seeds and compared field by field. Their numerical fields agreed exactly on the tested platform; cross-platform reproduction should still use the documented tolerance because floating-point libraries can differ.

We additionally test $N=5$ with Hamiltonian seed 93401 and input seed 271901. Let $W_0=ZZ^\dagger/\tr(ZZ^\dagger)$, where the entries of $Z$ are independent standard complex Gaussian draws. The coherent full-rank input is
\begin{equation}
\rho_0=0.55\ket\psi\bra\psi+0.45W_0,\qquad
\ket\psi=\sqrt{0.23}\ket0+\sqrt{0.77}\ket{v_R},
\end{equation}
where $v_R$ is a normalized Gaussian vector orthogonal to the target. The symbol $W_0$ here denotes a density matrix, not the Clifford circuit $W$. We also test the maximally mixed state, a pure zero-overlap state and the exact target. These cases include nonzero target--bright coherences and both endpoint values of the initial fidelity.

The uniform protocol uses $\gamma_i/J=1$ and $\eta_i=0.5$. The inhomogeneous protocol uses
\begin{align}
(\gamma_i/J)&=(0.35,0.70,1.10,1.40,2.00),\\
(\eta_i)&=(0.15,0.30,0.50,0.70,0.90).
\end{align}
The sufficient exponents are computed from the actual minimum and sum of $\eta_i\gamma_i$. For every input, we evaluate the full conditional density equation at 121 times from $Jt=0$ to 12. Density positivity, Hermiticity, target-weight conservation, and both success and fidelity bounds are checked throughout this grid. The largest bound violation or target-weight drift is below $5.6\times10^{-16}$. Such residuals are compared with an absolute tolerance of $2\times10^{-9}$ rather than treated as exact zeros.

An independent direct-matrix DOP853 integration uses relative tolerances $10^{-7},10^{-9},10^{-11}$ and absolute tolerance equal to one hundredth of each relative tolerance. The maximum Frobenius discrepancy from sparse exponential action decreases to below $6.6\times10^{-12}$ at the tightest setting. A dense exponential of the $1024\times1024$ Liouville matrix provides a third calculation at $Jt=0.35$: its mixed-state output agrees with sparse action within $1.5\times10^{-16}$. The corresponding Choi matrices have minimum eigenvalues no smaller than $-6.1\times10^{-15}$; the minimum eigenvalue of the trace-decrease slack is zero to reported precision. Choi reshuffling is first checked against the identity channel. These are numerical complete-positivity checks at the specified size and time, not a replacement for a general mathematical proof.

Finally, we compare the exact finite-step weak instrument with the continuous master equation. Each step first applies $e^{-iH\Delta t}$ and then all defect checks with $\kappa_i=\gamma_i\Delta t$. The retained check map is
\begin{equation}
\rho\longmapsto M_{i,0}\rho M_{i,0}^\dagger+(1-\eta_i)M_{i,1}\rho M_{i,1}^\dagger.
\end{equation}
Every finite step is a completely positive, trace-decreasing instrument. At $Jt=1.2$, using 16 through 1024 steps by successive doubling gives a density-error step-halving order tending to one (1.0020, 1.0010 and 1.0005 for the final three pairs). This validates the expected first-order discretization for this example; it does not equate a finite-step hardware sequence with its limiting differential equation. Under ideal uniform monitoring ($\gamma_i/J=1$, $\eta_i=1$), the physical-frame propagation of the coherent mixed input is also checked against conjugated syndrome-frame propagation at $Jt=1.2$, with Frobenius discrepancy below $2.2\times10^{-16}$. All new data and checks are generated by \texttt{supplementary\_validation.py}.

\begin{figure}[htbp]
\centering\includegraphics[width=\linewidth]{figS1_numerical_validation.pdf}
\caption{\textbf{Mixed-state filtering and independent numerical convergence checks.} \textbf{a}, Conditional fidelity of the coherent full-rank input for uniform and inhomogeneous monitoring. Solid curves are full density-matrix calculations; dashed and dotted curves are the analytic lower and upper bounds, respectively. \textbf{b}, No-click probability for four input classes under the inhomogeneous protocol, including exact zero and unit overlap. The zero-overlap state remains orthogonal to the target despite rare acceptance. \textbf{c}, Unnormalized density-matrix Frobenius error and normalized fidelity error of the exact finite-step instrument relative to continuous evolution at $Jt=1.2$. The slope-one line is a visual guide with arbitrary vertical placement, not a fitted result. \textbf{d}, Maximum Frobenius discrepancy over the time grid between direct-matrix DOP853 integration and sparse exponential action, as a function of the requested relative tolerance; the absolute tolerance is smaller by a factor of 100. All panels use $N=5$ and the explicitly specified seeds and rates.}
\label{fig:validation}
\end{figure}
\FloatBarrier

\section{Compatibility of overlapping product-basis windows}
Consider windows with starts $s_a$ and a repeated local Pauli string $p_0\ldots p_{R_w-1}$. They can all be obtained from one global product-basis setting if and only if
\begin{equation}
p_{x-s_a}=p_{x-s_b}\quad\text{for every shared site }x\in W_a\cap W_b.
\label{eq:windowcondition}
\end{equation}
Necessity follows because one physical qubit cannot simultaneously be measured in two distinct sharp Pauli bases on the same copy. Sufficiency follows by assigning the common required basis at every covered site. This is a simple consistency condition on all overlaps, not only on adjacent endpoints.

For regular starts separated by $R_w-1$, only one site overlaps and the condition reduces to equality of the first and last symbols. An additional right-anchored window can overlap several sites and introduces more constraints. Using $R_w=4$, stride three and $N=20$, the zero-based starts are $0,3,6,9,12,15,16$. In the final pair, the string XZYX requires $Z$ and $X$, respectively, at site 16. It cannot therefore be realized as one common product-basis setting. The strings YXYX and ZZZX already have unequal endpoints on regular stride-three overlaps. These are literal consistency issues for the stated window prescription in Ref.~\cite{Li2026Fingerprint}, not a reanalysis of its classification dataset.

\begin{table}[htbp]\centering
\caption{Exhaustive compatibility enumeration of all 81 four-site Pauli strings, repeated over the prescribed stride-three, right-anchored windows.}
\begin{tabular}{rrp{0.52\linewidth}}\toprule
$N$ & Compatible strings & Description\\\midrule
20 &3 &XXXX, YYYY, ZZZZ\\
30 &3 &XXXX, YYYY, ZZZZ\\
40 &27 &All strings with equal first and last symbol\\
60 &3 &XXXX, YYYY, ZZZZ\\
100 &27 &All strings with equal first and last symbol\\\bottomrule
\end{tabular}\end{table}

A correction can assign the final window the restriction of a globally fixed basis, or acquire incompatible windows in additional configurations. The local distributions then need to be associated with their actual measurement settings. Neither correction converts complete product readout into a nondemolition parity measurement. This is why the present scheme specifies a new instrument rather than transferring a classifier's input representation unchanged.

\section{Prior work, claim boundaries and submission status}
Exact scar construction is rooted in the Shiraishi--Mori framework and related algebraic or matrix-product-state constructions.\cite{ShiraishiMori2017,LinMotrunich2019,Moudgalya2018,Moudgalya2020MPS} Tower dynamics, quasiparticle interpretations and topological examples add structures absent from a single stationary cluster vector.\cite{Choi2019,SchecterIadecola2019,Mark2020,Ok2019,Chandran2023} Modern scar reviews describe these distinctions and the role of fragmentation.\cite{Serbyn2021,Moudgalya2022} Our family is an explicit instance of embedding, not a claim of a new universal construction.

There are particularly close antecedents. Stabilizer scars and cluster-state parent Hamiltonians are explicit in Refs.~\cite{StabilizerScars2025,Dooley2026}; transmon protocols include a single cluster scar.\cite{Larsen2026} Nondemolition stabilizer verification and preparation are established in Ref.~\cite{Liu2021QND}. Scars embedded into decoherence-free subspaces are treated in Ref.~\cite{Wang2024DFS}, and non-Hermitian stabilization is developed in Refs.~\cite{Chen2023NH,Omiya2025}. Scalable scar-based benchmarking is addressed in Ref.~\cite{Hartse2026Benchmark}. Accordingly, combining a cluster eigenstate, local annihilating jumps and conditional amplification does not by itself establish new physics.

The present exact statements are the finite-time identities, inequalities and operational distinctions derived above. They are presented as explicit consequences within this benchmark; no claim is made that these general mathematical tools are first introduced here. A publication-level novelty claim would require an additional result demonstrably beyond these antecedents, such as a proven task advantage under matched resources or a new experimentally tested mechanism. The available finite-size data and proofs do not establish such an advantage or guarantee a particular journal's editorial standard.

The single target has area-law entanglement, is an exact mid-spectrum eigenvector and is a local-observable outlier in the computed samples. Bulk level statistics approach the GUE reference over the accessible sequence. These are finite-size scar diagnostics. They do not prove asymptotic ETH, absence of all possible coefficient-dependent integrability, an open-system phase transition, a robust scar tower, protected encoded quantum information or fault tolerance. A thermal stabilizer counterexample\cite{Hokkyo2026} reinforces why stabilizer identity must not substitute for these diagnostics.

Finally, the manuscripts are anonymized research drafts arranged as Introduction, Results, Discussion, Methods, data/code statements and separate supplementary material. Author identities, affiliations, contributions, funding and conflicts require real author declarations before submission and are not invented. The journal's publicly accessible author guidance permits a readable compiled PDF for initial assessment and does not require imitation of its typeset production layout. This package therefore includes compiled PDFs and editable LaTeX; it is not a claim of editorial approval, peer review or acceptance.

\FloatBarrier
\bibliography{references}